\documentclass[12pt]{article}%
\pdfoutput=1
\usepackage[nosort]{cite}
\usepackage{graphicx}
\usepackage{subcaption}
\usepackage{multicol}
\usepackage{amsfonts}
\usepackage{amssymb}
\usepackage{amsmath}
\usepackage{heck}
\usepackage{setspace}
\usepackage{verbatim}
\usepackage{color}
\usepackage{longtable}
\usepackage{float}
\usepackage{tikz}
\usepackage[margin=1in]{geometry}
\usepackage{titletoc}%
\providecommand{\U}[1]{\protect\rule{.1in}{.1in}}
\pdfoutput=1
\usetikzlibrary{decorations.markings}

\numberwithin{equation}{section}

\newcommand{\ba}{\begin{eqnarray}}
\newcommand{\ea}{\end{eqnarray}}

\newcommand{\decoup}{\sqcup}
\begin{document}

\date{May 2015}

\title{Geometry of 6D RG Flows}

\institution{UNC}{\centerline{${}^{1}$Department of Physics, University of North Carolina, Chapel Hill, NC 27599, USA}}

\institution{UCSBmath}{\centerline{${}^{2}$Department of Mathematics, University of California Santa Barbara, CA 93106, USA}}

\institution{UCSBphys}{\centerline{${}^{3}$Department of Physics, University of California Santa Barbara, CA 93106, USA}}

\institution{HARVARD}{\centerline{${}^{4}$Jefferson Physical Laboratory, Harvard University, Cambridge, MA 02138, USA}}

\authors{Jonathan J. Heckman\worksat{\UNC}\footnote{e-mail: {\tt jheckman@email.unc.edu}},
David R. Morrison\worksat{\UCSBmath, \UCSBphys}\footnote{e-mail: {\tt drm@physics.ucsb.edu}},\\[2mm]
Tom Rudelius\worksat{\HARVARD}\footnote{e-mail: {\tt rudelius@physics.harvard.edu}},
and Cumrun Vafa\worksat{\HARVARD}\footnote{e-mail: {\tt vafa@physics.harvard.edu}}}

\abstract{We study renormalization group flows between six-dimensional superconformal field theories (SCFTs) using
their geometric realizations as singular limits of F-theory compactified on elliptically fibered Calabi-Yau threefolds.
There are two general types of flows:  One corresponds to giving expectation values to scalars in the tensor
multiplets (tensor branch flow) realized as resolving the base of the geometry.  The other corresponds to giving
expectation values to hypermultiplets (Higgs branch flow)
realized as  complex structure deformations of the geometry.
 To corroborate this physical picture we calculate the change in the anomaly polynomial for these theories,
finding strong evidence for a flow from a UV fixed point to an IR fixed point.  Moreover, we find evidence
against non-trivial dualities for 6D SCFTs.  In addition we find non-trivial RG flows for theories realizing
small $E_8$ instantons on ALE spaces. }

\maketitle

\tableofcontents

\enlargethispage{\baselineskip}

\setcounter{tocdepth}{2}

\newpage

\section{Introduction \label{sec:INTRO}}

One of the central concepts in the study of quantum field theory is
renormalization group (RG) flows between different theories. This is
especially important in the context of conformal field theories (CFTs) which
correspond to fixed points of the renormalization group equations. Given two
CFTs, a natural question to ask is whether there is a corresponding RG flow
which connects two theories. In general terms, these flows are triggered
either by an operator perturbation of the original UV CFT, or by a non-zero
operator vev.

In this note we study the question of RG flows for six-dimensional SCFTs, i.e
those with at least eight real supercharges. The best evidence for the
existence of these theories comes from string theory \cite{Witten:1995ex,
Witten:1995zh, Strominger:1995ac, WittenSmall,
Ganor:1996mu,MorrisonVafaII,Seiberg:1996vs, Seiberg:1996qx, Bershadsky:1996nu,
Brunner:1997gf, Blum:1997fw, Aspinwall:1997ye, Intriligator:1997dh,
Hanany:1997gh}, and recently a classification of 6D\ SCFTs which can be
generated by string compactification has been achieved \cite{Heckman:2013pva,
Gaiotto:2014lca, DelZotto:2014hpa, Heckman:2014qba, DelZotto:2014fia,
Heckman:2015bfa, Bhardwaj:2015xxa}. Given this list of theories, it is natural
to ask about RG flows which can potentially connect these theories together.

The classification results of reference \cite{Heckman:2015bfa} rely on the
F-theory realization of these theories. To make an SCFT, we start with
F-theory on an elliptically fibered Calabi-Yau threefold $X$ with the base $B$
a non-compact K\"ahler surface. By simultaneously contracting $\mathbb{P}^{1}$'s
of this geometry, we reach a conformal fixed point. Reference
\cite{Heckman:2015bfa} determined all possible configurations of curves which
can simultaneously contract, and moreover, all possible ways to generate an
elliptic fibration over a given base.

The geometric characterization also provides insight into the moduli space of
these theories. Indeed, starting from a smooth Calabi-Yau geometry, we reach a
conformal fixed point by taking a singular limit in the Calabi-Yau moduli
space. At a smooth point in the moduli space, the basic moduli of our theory
organize according to scalars in tensor multiplets and scalars in
hypermultiplets. The geometric interpretation of the tensor multiplet scalars
is simply the volumes of the $\mathbb{P}^{1}$'s of the base. Complex structure
moduli correspond to complex scalars of the effective theory, and as such fill
out half of the scalar degrees of freedom of a $(1,0)$ hypermultiplet in six
dimensions.\footnote{The scalars of the hypermultiplet moduli are captured by
the complex structure moduli, as well as the intermediate Jacobian of the
Calabi-Yau threefold.\ Owing to the fact that there is an $SU(2)$ R-symmetry
which acts on the tangent space to the hypermultiplet moduli space, we shall
often simply speak of the complex structure moduli as indicating the presence
of hypermultiplet moduli. This observation turns out to be quite important in
the analysis of the limiting behavior of T-brane solutions
\cite{Anderson:2013rka}.} Going to a singular point in the geometric moduli
space corresponds to proceeding to the UV.

To study flows between SCFTs, we can therefore start with a singular geometry
(i.e.\ a UV SCFT), and then activate vevs for scalar operators of the theory.
This triggers a flow to a new theory, leading us to a geometry which may still
be singular (and thus an IR SCFT). Our task therefore reduces to correctly
identifying which deformations of the geometry actually correspond to operator
vevs.

In order to corroborate our picture of flows between theories, we perform
various checks on the match with geometry. First of all, we show that the
resulting moduli are associated with normalizable modes of the effective field
theory. Additionally, we need to be able to quantitatively compare the number
of degrees of freedom present in a 6D\ SCFT. Along these lines, we use the
recently developed methods of references \cite{Ohmori:2014pca, Ohmori:2014kda}
to compute anomaly polynomials for 6D\ SCFTs to compare candidate SCFTs. A
specific difference in the value of the anomaly polynomial, consistent with 't Hooft anomaly matching
conditions \cite{Ohmori:2014kda,Intriligator:2014eaa}   is a strong indication that
there is an RG\ flow. Since we can also typically identify which
theory is associated with a more singular point in moduli space, we also have
a clear indication about how we can flow from the UV to the IR.

Using anomaly polynomials in the context of a number of examples leads us
to a surprising conclusion that there apparently are no dualities for $(1,0)$ SCFTs.
In particular we show that candidate dual theories have distinct anomaly polynomials
ruling them out as duals of one another.   This observation suggests that the classification of
\cite{Heckman:2015bfa} does not need to be modded out by duality equivalences.  We also
study a class of models  from reference \cite{Heckman:2015bfa}, associated with the dual
F-theory description of heterotic instantons probing an ADE\ singularity.
For a fixed choice of boundary data for the heterotic instanton, we find that
there are a number of dual F-theory models. We present an explicit hierarchy of flows between
theories with given boundary data. The full moduli space therefore assembles according to a partially
ordered set dictated by RG\ flows and we find a {\it single} class of theories
related by RG\ flows for each class of asymptotic bundle data for the heterotic string.

The rest of this note is organized as follows. First in section
\ref{sec:FLOWS} we provide some more details on the general expected
correspondence between deformations of an F-theory geometry and the
associated flows in moduli space. We then turn in section \ref{sec:APOLY}
to quantitative checks on this picture, focussing on the role of the anomaly
polynomial for a 6D\ SCFT, and how it tracks details of flows between
theories. In section \ref{sec:EXAMPLES} we give several illustrative examples of both tensor branch
and Higgs branch flows. We conclude in section \ref{sec:CONC}.

\section{Geometry of 6D\ RG\ Flows \label{sec:FLOWS}}

In this section we discuss in more detail the sense in which deformations in
the geometry of an F-theory compactification can be translated into vevs of
operators in a 6D\ SCFT. To frame our discussion we shall often consider
F-theory on an elliptically fibered Calabi-Yau threefold $X\rightarrow B$. To
make an SCFT, it suffices to consider $B$ a non-compact K\"{a}hler surface
such that all rational curves are simultaneously contractible in the
K\"{a}hler cone. The SCFT\ point corresponds to this singular limit where the
volumes of all curves have shrunk to zero size. The data specifying
an F-theory compactification is given by sections
$f$ and $g$ of $\mathcal{O}_{B}(-4K_{B})$ and $\mathcal{O}_{B}(-6K_{B})$, respectively, leading
to an associated Calabi--Yau threefold whose (minimal)
Weierstrass form  is:%
\begin{equation}
y^{2}=x^{3}+fx+g. \label{minWeier}%
\end{equation}

In the special case of an SCFT, the minimal Weierstrass equation will
typically be quite singular. To make sense of the physical model, we
 consider blowups of the geometry  and/or smoothing
deformations to reach a less singular geometry. For example, if we have a
collection of curves in the base where the Kodaira-Tate fiber becomes too
singular, we will need to perform blowups in the base. If no finite sequence
of blowups is available, we will discard the putative F-theory model.

Since being able to reach a smooth geometry ensures  the existence of
an F-theory model, we can also choose to reverse the above
procedure:\ We can start from a smooth Calabi-Yau $\widetilde{X}$ and then
take a degeneration limit either in K\"{a}hler moduli or complex structure
moduli. In this sense, we can consider various intermediate stages before
reaching the maximally singular case of equation (\ref{minWeier}). For
example, we can work over a smooth base, but allow degenerations in the
elliptic fiber, or conversely, we can consider a singular base, but with a
smooth fiber. In some cases, the data of the base and fiber are correlated and
cannot be disentangled.

Suppose then that we are working with an F-theory model given by a smooth base
$B$, and an associated elliptic fibration with smooth total space and all
fibers having dimension one.  There is then a
well-defined prescription for  extracting the 6D effective
field theory for this system. The field content will consist of some number of
tensor multiplets, hypermultiplets, and vector multiplets. The K\"ahler
parameters of the base $B$ correspond to background vevs for the real scalars
of the tensor multiplet, and the complex structure moduli correspond to half
of the vevs for the hypermultiplets.

More precisely, we need to know which of the geometric deformations actually
translate to physical parameters of our system. These are described by the
normalizable deformations, namely those with finite non-zero kinetic terms at
a smooth point of the moduli space. Letting $\delta J_{(1,1)}$ denote
variations of the K\"ahler form and $\delta\Omega_{(2,1)}$ denote the variations
of the holomorphic three-form, we need to ensure that:%
\begin{equation}
\int_{B}\delta J_{(1,1)}\wedge\delta J_{(1,1)}<\infty\text{ \ \ and \ \ }%
\int_{\widetilde{X}}\delta\Omega_{(2,1)}\wedge\delta\overline{\Omega}_{(1,2)}<\infty.
\end{equation}
In the latter case, note that we also need to pick a resolution $\widetilde{X} \rightarrow X$ 
of the singular Calabi-Yau $X$ to carry out the integral.

Such normalizable modes specify deformations of the effective field theory. By
tuning their values, we can reach singular points
of the moduli space, i.e.\ where we have a 6D SCFT.
A geometric criterion for identifying which complex structure limits of
Calabi--Yau threefolds are at finite distance in the moduli space (and
hence represent normalizable modes) was found in \cite{Hayakawa,Wang}:
the singularities must be so-called ``canonical singularities.''
Moreover, the techniques of \cite{Grassi91} can be used to show that
these are the same singularities which can be resolved after blowing
up the base of an elliptic fibration.

Of course, in physical terms we start at the most singular geometries and then
proceed to activate vevs for operators to flow to new IR\ fixed points. A common feature of such flows is that starting from a
UV fixed point, a flow to the infrared can sometimes lead to completely decoupled CFTs in the IR:
\begin{equation}
CFT_{UV} \rightarrow CFT^{(1)}_{IR} \decoup \cdots \decoup CFT^{(m)}_{IR}.
\end{equation}
(The ``disjoint union'' symbol ${\decoup}$ is used to denote decoupled theories.)
When the context is clear, we shall sometimes omit these additional decoupled sectors. However, to properly match all UV and IR anomalies,
we must include all such sectors.

In geometric terms, there are two ways for us to trigger a flow, corresponding to K\"ahler deformations and complex structure deformations (i.e.\ smoothings). Physically, the K\"ahler deformations trigger tensor branch flows and the complex structure deformations trigger Higgs branch flows. We can also have various mixed branch flows obtained by combining these two basic operations. Indeed, these two branches of moduli space can 
intersect along lower-dimensional subspaces.

The case of activating K\"ahler parameters is by now well-known, and corresponds to
giving tension to at least some subset of strings of a theory. For example, we
can flow between the different $A_{N}$ $(2,0)$ theories by starting with a
configuration of $N$ collapsed $-2$ curves in a base $B$:%
\begin{equation}
A_{N}\text{ }(2,0)\text{ theory: }\underset{N}{\underbrace{2,...,2}}%
\end{equation}
Resolving these curves from left to right in the associated Dynkin diagram
triggers a sequence of flows:%
\begin{equation}
A_{N}\overset{RG}{\rightarrow}A_{N-1} \decoup T_{(2,0)}\overset{RG}{\rightarrow}%
...\overset{RG}{\rightarrow}A_{1} \decoup (T_{(2,0)})^{\decoup (N - 1)},
\end{equation}
where at each stage of this resolution process, we have some
additional decoupled $(2,0)$ tensor multiplets, each of which we denote
as $T_{(2,0)}$. In the M-theory description, we are pulling
apart a stack of $N$ M5-branes one at a time. In this sense,
the $A_{N}$ theory is in the UV and the $A_{1}$ theory with $(N-1)$
decoupled $(2,0)$ tensor multiplets is in the IR.

As another class of examples, we can consider starting from the rank
$N$ E-string theory:%
\begin{equation}
\text{rank }N\text{ E-string theory: }\underset{N}{\underbrace{1,2,...,2}}.
\end{equation}
Note that here we can either flow to a lower rank E-string theory (i.e.\ by
resolving the $-2$ curve on the right to finite size), or we can flow to an
$A$-type $(2,0)$ theory. Again, the M-theory description is helpful in tracking the various decoupled SCFTs in the infrared.

Now, in addition to motion in the K\"{a}hler moduli space, we can also study
complex structure deformations by activating vevs for the operators
parameterizing the Higgs branch. Some of these cases have a clear
interpretation in terms of a weakly coupled Lagrangian description. For
example, consider the local model for the collision of a non-compact curve of
$A_{N}$ singularities with a non-compact curve of $A_{M}$ singularities.
Letting $u=0$ and $v=0$ denote the corresponding loci, a local model for the
Calabi-Yau threefold is:%
\begin{equation}
y^{2}=x^{2}+u^{N}v^{M}\text{.}%
\end{equation}
In physical terms, there is a collection of $N\times M$ bifundamental
hypermultiplets trapped at the intersection point $u=v=0$. We can move onto
the Higgs branch by activating a vev for these hypermultiplets. Assuming
without loss of generality that $N\leq M$, the deformed geometry takes the
form \cite{BHVI}:%
\begin{equation}
y^{2}=x^{2}+v^{M-N}\underset{i=1}{\overset{N}{%
{\displaystyle\prod}
}}(uv-t_{i})\text{,}%
\end{equation}
i.e.\ we break $SU(N)\times SU(M)$ to $SU(M-N)$. The parameters $t_{i}$ are
specified by the vevs of the hypermultiplets. More precisely, we can, by a
choice of appropriate flavor rotation view the hypermultiplets as transforming
in the adjoint representation of the diagonal $SU(N)_{\text{diag}}\subset
SU(N)\times SU(M)$. The unfolding parameters for the corresponding singularity
then follow the standard rules for geometric engineering.

We can also produce an interacting fixed point by gauging some of our symmetries.
For example, to gauge the $SU(N)$ subgroup of the $SU(N)\times SU(2N)$
global symmetry of a theory with
$N\times 2N$ bifundamental hypermultiplets, we can
consider an $SU(N)$ gauge theory on a curve of self-intersection $-2$. Such a
theory must include $2N$ flavors in order to satisfy the anomaly cancellation conditions for the
gauge symmetry, so it has a flavor symmetry of $SU(2N)$ (while the $SU(N)$
factor has been gauged).

We can now move onto a partial Higgs branch of this theory, breaking the
gauge symmetry to $SU(N-1)$. The resulting theory consists of the gauge theory sector, as well as $2N - 1$ hypermultiplets which are
singlets under this gauge symmetry. The full Higgs branch flow is then:
\begin{equation}
[SU(2N)] SU(N) \overset{RG}{\rightarrow} [SU(2N - 2)] SU(N-1) \decoup (2N - 1) \, \mathrm{singlets},
\end{equation}
where here, the  group in square brackets denotes the flavor group.

More generally, however, there need not exist a weakly coupled interpretation
of these operator vevs. To give examples along these lines, consider the case
of a conformal matter sector (in the sense of \cite{DelZotto:2014hpa, Heckman:2014qba})
localized at the intersection of two $E_{8}$
seven-branes respectively located at $u=0$ and $v=0$:%
\begin{equation}
y^{2}=x^{3}+u^{5}v^{5}\text{.}%
\end{equation}
There is conformal matter localized at $u=v=0$. From this singular model, we
can either move onto the tensor branch, or onto the Higgs branch. Moving onto
the tensor branch amounts to introducing a sequence of blowups of the singular
point of the base. We must continue to blowup until the elliptic fiber over
each blown up curve is in Kodaira-Tate form. As first derived in
reference \cite{Aspinwall:1997ye},
this leads to a sequence of eleven blowups, leading to the
geometry:%
\begin{equation}
\lbrack E_{8}]\oplus\lbrack E_{8}]\simeq\lbrack E_{8}%
]1,2,2,3,1,5,1,3,2,2,1[E_{8}]\text{,}%
\end{equation}
where on the left side we have indicated the conformal matter by the symbol
$\oplus$, and on the righthand side we have indicated the explicit resolution
of curves. The corresponding eleven volume moduli parameterize the tensor
branch of the SCFT. One set of flows corresponds to having some subset of the
volume moduli go to non-zero values. For example, we can initiate a flow by resolving the middle $-5$ curve. When we do this, the two sequences of
curves $1,2,2,3,1$ and $1,3,2,2,1$ decouple. Additionally, the theory on the $-5$ curve decouples as another SCFT. This sector consists of an additional $(1,0)$ tensor multiplet, and a gauge theory with gauge group $F_4$. Now, an important feature of this gauge theory is that in the infrared, the value of its gauge coupling is formally zero, the reason being that the tensor multiplet scalar has picked up a formally infinite vev. This geometry corresponds, in the M-theory
picture as having an M5-brane probing an $E_8$ singularity \cite{DelZotto:2014hpa} and the above
tensor branches emerge from fractionating the M5-brane to 12 (or fewer) pieces.

\begin{figure}
\begin{center}
\includegraphics[trim=25mm 30mm 30mm 10mm, clip, width=60mm]{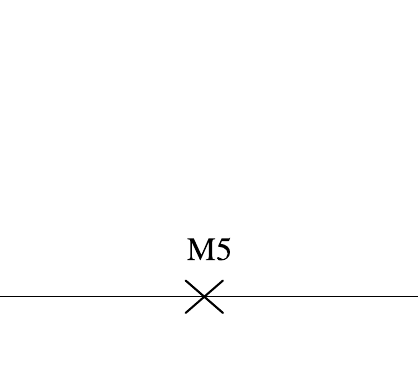}
\includegraphics[trim=25mm 30mm 30mm 10mm, clip, width=60mm]{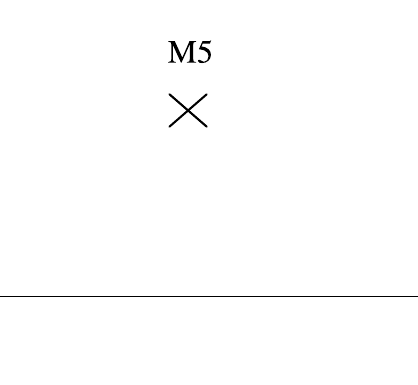}
\caption{M-theory realization of the flow from $E_8 \oplus E_8$ conformal matter to a single tensor multiplet.  There is an M5-brane probing the $E_8$ singularity in the UV (left), which lifts off the singularity in the IR (right).}
\label{M5lifting}
\end{center}
\end{figure}

Instead of considering a flow on the tensor branch, we can also consider
complex deformations. For example, we can consider the partial unfolding to
the model:%
\begin{equation}
y^{2}=x^{3}+\left(  uv - \beta\right)  ^{5}.
\end{equation}
The parameter $\beta$ is our complex structure modulus, and it corresponds to
a brane recombination mode \cite{DelZotto:2014hpa, Heckman:2014qba}. Indeed, for generic $\beta$, we have a single
$E_{8}$ seven-brane located at $uv = \beta$. This generalizes the usual notion
of Higgsing and brane recombination to the case of a conformal matter sector.  In the M-theory
picture (see Figure \ref{M5lifting}) this corresponds to lifting off the M5-brane from the $E_8$ singularity.
The theory flows in the IR to a free theory of a single $(2,0)$ tensor multiplet, i.e.\ the theory of an M5-brane away from the singularity.

\begin{figure}[ptb]
\begin{center}
\begin{subfigure}[a]{0.3\textwidth}
                \includegraphics[trim=10mm 10mm 10mm 10mm, clip, width=50mm]{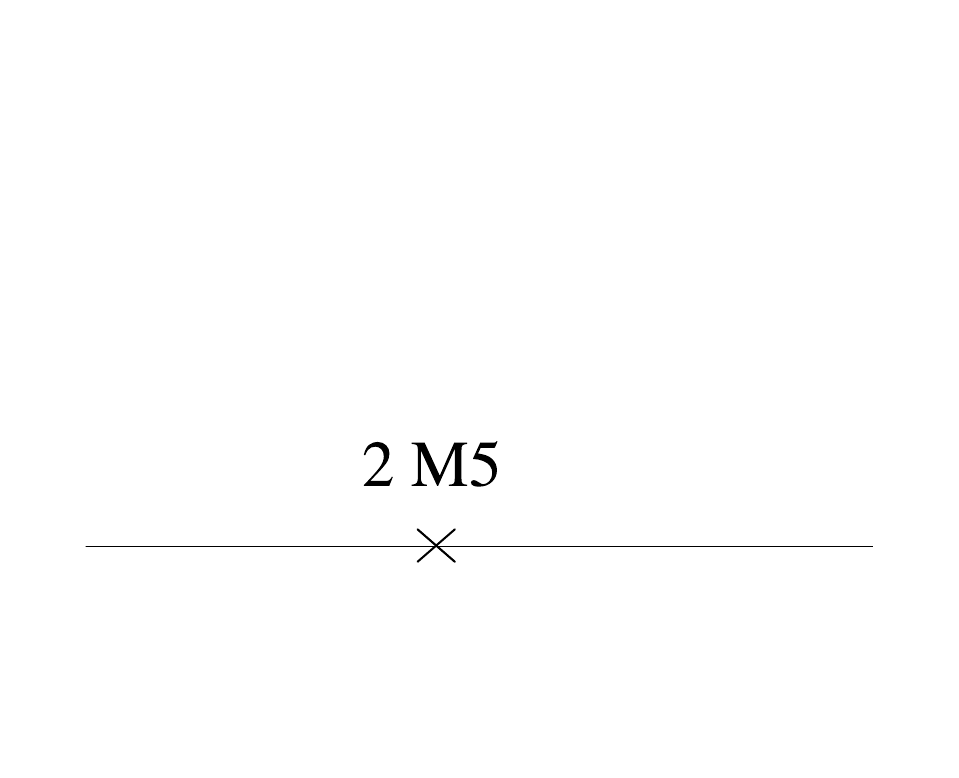}
                \caption{Two M5-branes on an $E_8$ singularity.}
                \label{twoM5on}
\end{subfigure}
\begin{subfigure}[b]{0.3\textwidth}
                \includegraphics[trim=10mm 10mm 10mm 10mm, clip, width=50mm]{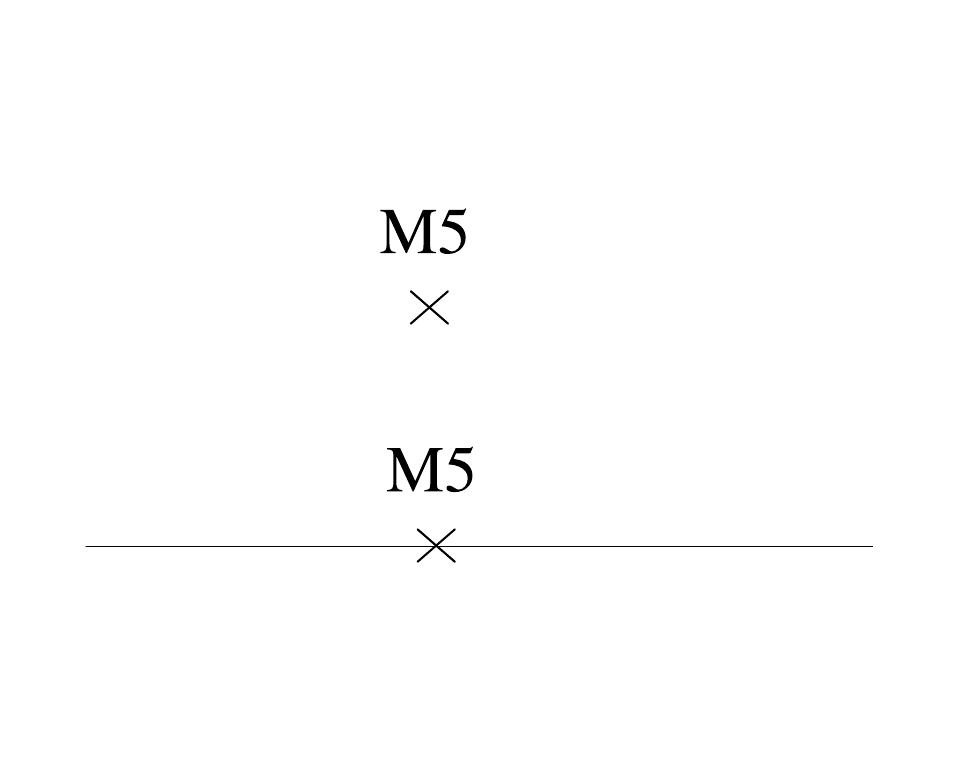}
                \caption{One M5-brane lifted.}
                \label{twoM5oneoff}
\end{subfigure}
\begin{subfigure}[c]{0.3\textwidth}
                \includegraphics[trim=10mm 10mm 10mm 10mm, clip, width=50mm]{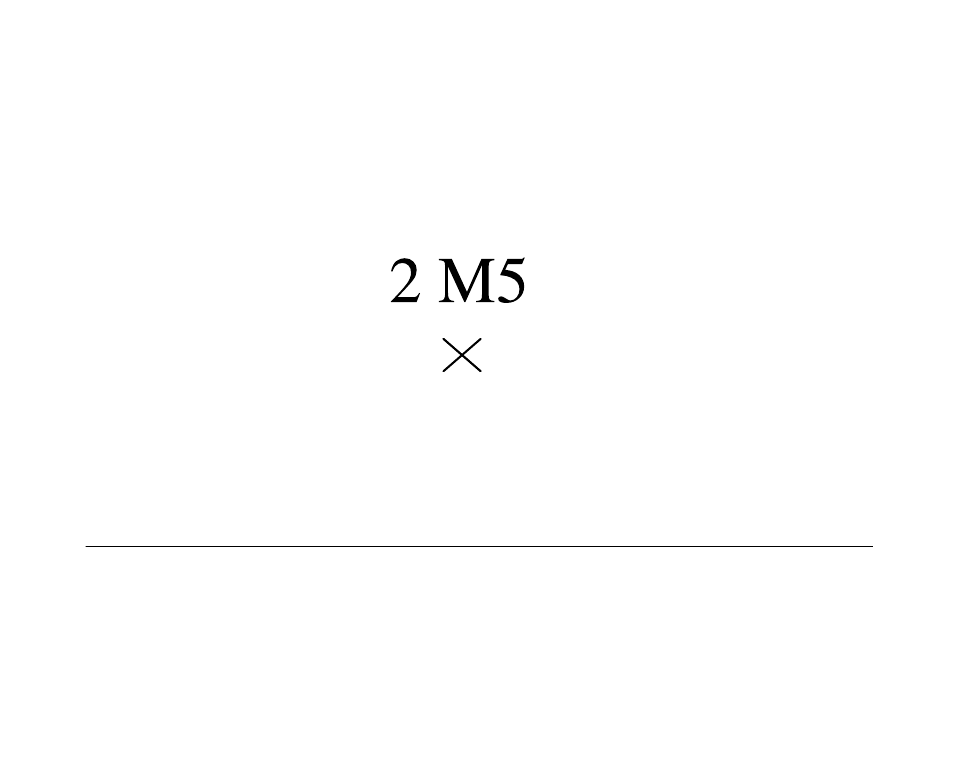}
                \caption{Two M5-branes lifted.}
                \label{twoM5twooff}
\end{subfigure}
\caption{Two M5-branes probing an $E_8$ singularity.  The theory flows from (a) to (b) to (c).}
\end{center}
\end{figure}

Similar considerations also hold for the generalized quiver theories of the
type encountered in \cite{DelZotto:2014hpa, Heckman:2014qba, Heckman:2015bfa}.
For example, consider the case of a $-2$ curve which supports an $\mathfrak{e}_{8}$ fiber, which in turn
intersects two non-compact $E_{8}$ seven-branes according to the diagram:%
\begin{equation}
\lbrack E_{8}] \oplus \overset{e_{8}}{2} \oplus [E_{8}];
\end{equation}
the corresponding M-theory picture is given by having two M5-branes probing the $E_8$ singularity, (see Figure \ref{twoM5on}).
This leads to two $(E_{8},E_{8})$ conformal matter sectors, one for each
intersection of the $-2$ curve with a non-compact seven-brane. If we expand the middle $-2$ curve with $\mathfrak{e}_8$ gauge symmetry
to large size, we get a flow on the tensor branch to two decoupled $(E_8 , E_8)$ conformal matter systems, a $(1,0)$ tensor multiplet, and
$248$ free vector multiplets. Alternatively, we can flow on the Higgs branch by giving a vev to
one such conformal matter sector. This initiates an unfolding, i.e.\ breaking
pattern:%
\begin{equation}
\lbrack E_{8}]\oplus\overset{\mathfrak{e}_{8}}{2}\oplus\lbrack E_{8}]\rightarrow(\lbrack
E_{8}]\oplus\lbrack E_{8}]) \decoup T_{(2,0)},
\end{equation}
which in the M-theory picture corresponds to lifting one M5-brane off the singularity (see Figure \ref{twoM5oneoff}).
Physically, activating a vev for the conformal matter initiates a breaking
pattern to the diagonal $E_{8}$ subgroup of $[E_{8}]\times \mathfrak{e}_{8}$. As one of
the groups participating in the Higgsing is a flavor symmetry, the unbroken
symmetry group is also ungauged (i.e.\ at arbitrarily small gauge coupling).  One can also
lift both M5-branes together, which makes the theory flow to the $A_1$ (2,0) theory
(see Figure \ref{twoM5twooff}).

\section{Anomaly\ Polynomial\ Analysis \label{sec:APOLY}}

In the previous section we presented a general description of possible flows
between 6D theories, using the geometry as a guide to possible operator vevs,
i.e.\ flows on the tensor branch and/or generalized Higgs branch. In general,
however, it can be mathematically challenging to determine when a given
complex structure deformation corresponds to a normalizable mode of the
physical theory. Indeed, precisely because our defining geometries are so
singular, specifying what counts as a \textquotedblleft lower order
deformation\textquotedblright\ which can initiate a flow is a rather subtle question.

To complement this perspective, we can also study the anomalies of global
symmetry currents of our system. A change in the anomaly polynomial after applying a
geometric deformation is a strong indication that this deformation has
initiated a flow to a new theory. Turning the discussion around, the analysis
of anomalies can provide a good check on candidate normalizable deformations
of a singular, non-compact model. In other words, comparison of anomalies gives us a
highly non-trivial tool to assess whether two putative theories are related by an RG flow.

The anomalies of a 6D\ SCFT are conveniently captured in terms of the anomaly
polynomial, i.e., a formal eight-form constructed from the background field
strengths of the system. Since all gauge anomalies must cancel to have a
consistent quantum theory, the main utility in the context of checking flows
will center on the global symmetry currents. This includes the 6D spacetime
diffeomorphisms associated with the tangent bundle $T$, the $SU(2)$ R-symmetry
which we associate with curvature $R$ in a background bundle, as well as curvatures $F_i$ in background bundles for any flavor symmetries which
an SCFT may possess.\footnote{A word of caution to the reader. In the
literature for supergravity models generated by F-theory compactifications, it
is common to let $R$ refer to the \textquotedblleft tangent bundle
contributions\textquotedblright\ in the 6D effective theory, and to completely
omit the contribution from the R-symmetry currents. The reason for this
omission is simply that in all of these models gravity is dynamical and there
is no $SU(2)$ R-symmetry.} The general structure of an anomaly
polynomial will then be built out of the formal characteristic classes
$p_{1}(T)$, $p_{2}(T)$, $c_{2}(R)$ and $c_{2}(F_{i})$, where the $p$'s denote
Pontryagin classes (i.e.\ $p_{1}$ is a four-form and $p_{2}$ is an eight-form)
and the $c_{2}$'s denote second Chern classes (i.e.\ four-forms). Given an
SCFT, we therefore get the anomaly polynomial:
\begin{align}
\mathcal{I}_{SCFT}  &  =\alpha c_{2}(R)^{2}+\beta p_{1}(T)^{2}+\gamma
c_{2}(R)p_{1}(T)+\delta p_{2}(T)\\
&  +\underset{i}{\sum}\delta_{i}c_{2}(R)c_{2}(F_{i})+\underset{i}{\sum}%
\gamma_{i}p_{1}(T)c_{2}(F_{i})+\underset{i}{\sum}\kappa_{i}\delta_{i}%
c_{2}(F_{i})c_{2}(F_{i}),
\end{align}
where in our conventions $c_2(F) = \frac{1}{4} \mathrm{Tr} F^2$, with $F$ the field strength.
The terms of the first line are common to all 6D\ SCFTs. The terms of the
second line depend on the details of the flavor symmetries present in a given theory.

In fact, even more is required in the difference $\mathcal{I}_{UV} - \mathcal{I}_{IR}$ of the anomaly polynomials for
UV and IR SCFTs. If we restrict our analysis to symmetries
which are common to the UV and IR theories, then the difference of the anomaly polynomials needs to be
a perfect square \cite{Intriligator:2014eaa} (by which we mean a quadratic form). The
reason is that 't Hooft anomaly matching for our theory can only proceed
if there is an appropriate Green-Schwarz term available to compensate for this difference. This in turn requires factorization of
the difference of the UV and IR anomaly polynomials into a perfect square. Since diffeomorphism invariance is always present, $p_1(T)$ also needs to appear in a perfect square. Additionally, since $p_2(T)$ can never factor to a perfect square, its coefficient in the anomaly polynomial is the same in the UV and the IR. Further, if the R-symmetry is unbroken by a flow, then $c_2(R)$ needs to appear in a perfect square.

Of course, in practice verifying the factorization condition for unbroken symmetries can turn out to be subtle. For example, when we initiate a flow to the infrared, we need to account for all possible decoupled SCFTs. Additionally, much as in the case of 4D SCFTs \cite{Intriligator:2003jj},
it is a priori possible that the infrared R-symmetry may mix with additional $SU(2)$ flavor symmetries. Further, such flavor symmetries may only emerge in the infrared. When we turn to examples of flows in section \ref{sec:EXAMPLES} we will indeed encounter various examples with emergent symmetries in the infrared.

The anomaly polynomial for all 6D\ SCFTs constructed via F-theory can
be determined using the methods developed in references \cite{Ohmori:2014pca,
Ohmori:2014kda}. Since we will be making heavy use of this result when we turn
to examples, we shall spell out some additional details about how to
algorithmically generate the anomaly polynomial of a 6D\ SCFT given an
explicit F-theory model.

The basic philosophy of this approach hinges on two observations. First of
all, since the real scalar of a tensor multiplet is neutral under the $SU(2)$
R-symmetry, we can pass onto the tensor branch. By anomaly matching
considerations, the anomaly polynomial on the tensor branch must match to that
of the SCFT.\footnote{In analogy with the case of lower-dimensional examples,
this of course assumes that there is no emergent $SU(2)$ flavor symmetry which
mixes with the putative R-symmetry.} The second observation is that in the
special case where the number of tensor multiplets $n_{\text{tensors}}$ is
equal to the number of simple gauge group factors $n_{\text{groups}}$,
compatability with the Green-Schwarz mechanism for the gauge group factors
imposes a unique structure for the resulting anomaly polynomial.

In more detail, the anomaly polynomial has a one-loop contribution, as well as
a contribution from the exchange of a tensor field. These two pieces combine
as:%
\begin{equation}
\mathcal{I}_{\text{tot}}= \mathcal{I}_{\text{1-loop}}+\mathcal{I}_{GS}.
\end{equation}
For anomaly cancellation to work via the Green-Schwarz mechanism, we must
demand that the one-loop contribution, to the anomaly polynomial factorizes
for all of the gauge groups. For the flavor symmetry factors, however, no such
factorization is necessary. So, in the special case where the number of tensor
multiplets is the same as the number of simple gauge group factors, we can
extract the total contribution to the anomaly polynomial by completing the
square of the one-loop contribution.

In general, however, anomaly cancellation in a 6D\ SCFT only imposes the
weaker condition that $n_{\text{tensors}}\geq n_{\text{groups}}$. When we have
more tensors than simple gauge groups, factorization alone does not dictate a
unique answer. Examples of this type include the E-string theory, as well as
theories containing E-strings interposed between simple gauge group factors:%
\begin{align}
\text{E-string Theory}  &  \text{: }1,2...,2\\
\text{Example Theory}  &  \text{: }(\overset{\mathfrak{e}_{8}}{12}%
),1,\overset{II}{2},\overset{\mathfrak{sp}_{1}}{2},\overset{\mathfrak{g}%
_{2}}{3}, \label{example theory}%
\end{align}
where for the convenience of the reader we have indicated the gauge group /
fiber type associated with each theory. Another class of examples are those
which contain just $-2$ curves interposed between gauge group factors.
Examples of this type include fiber enhancemnets over a chain of $-2$ curves:%
\begin{align}
A_{3}\text{ }(2,0)\text{ Theory: }  &  2,2,2\\
\text{Fiber Enhanced Theory: }  &  [SU(n)]\overset{\mathfrak{su}_{n}%
}{2},\overset{\mathfrak{su}_{n}}{2},\overset{\mathfrak{su}_{n}}{2}[SU(n)].
\end{align}

On the other hand, the anomaly polynomials for the $A_{k}$ $(2,0)$ theories
have been determined in references \cite{Duff:1995wd, Freed:1998tg,
Harvey:1998bx}, and the rank $k$ E-string theory anomaly polynomial has been
determined in reference \cite{Ohmori:2014pca}. This turns out to be enough to
fix the form of the anomaly polynomial for all SCFTs which arise in F-theory.
The main idea is that we interpret these theories without a simple gauge group
factor as a generalized type of matter which is interposed between two gauge
groups. In the example of line , we view the configuration $1,2$ as a
generalized matter of both an $\mathfrak{e}_{8}$ gauge theory (the $-12$
curve) and an $\mathfrak{sp}_{1}$ gauge theory (the rightmost $-2$ curve).
Viewing all such contributions as generalized matter, we have a system where
the remaining number of tensor multiplets matches to the number of simple
gauge group factors.

Now, an important result from reference \cite{Heckman:2015bfa} is that all
6D\ SCFTs which arise in F-theory are of precisely the form which can be
handled by the method of \cite{Ohmori:2014kda}. This means that to extract
this data, it is enough for us to carefully take into account the
contributions from these generalized matter contributions. Indeed, proceeding
inductively, one can start building up larger structures which must also work
in the same way. For example, the $(E_8,E_8)$ conformal matter of references
\cite{Aspinwall:1997ye,DelZotto:2014hpa, Heckman:2014qba} is a sequence of curves:%
\begin{equation}
\lbrack E_{8}],1,2,2,3,1,5,1,3,2,2,1,[E_{8}].
\end{equation}
So if our only concern is the contribution to the anomaly polynomial from the
global symmetry currents, we can build up to larger structures in one shot rather than adding
up the individual E-string theories and imposing factorization of the anomaly
polynomial iteratively.

It is therefore enough for us to fix the contribution from these
generalized types of matter, i.e., the contributions from the $A$-type $(2,0)$
theories and the rank $k$ E-strings which attach to simple gauge group
factors.\footnote{We thank K. Yonekura for helpful discussions on this point.}
The main subtlety we encounter in this procedure is the proper way to deal
with the center of mass degrees of freedom for these SCFTs once we couple them to larger structures.

For example, in the case of the $(2,0)$ theory realized by $k$ M5-branes in flat space, there is an
overall \textquotedblleft center of mass degree of freedom\textquotedblright%
\ described by a free $(2,0)$ tensor multiplet. In terms of $(1,0)$
multiplets, this is is a single hypermultiplet and a $(1,0)$ tensor
multiplet. By a similar token, when we consider the
anomaly polynomial for a rank $k$ E-string theory, there is a
hypermultiplet parameterizing the center of mass degree of freedom for the
$k$ small instantons inside an $E_{8}$ wall. In F-theory language, this additional hypermultiplet corresponds to the
point in the base where all the curves contract. Indeed, the hypermultiplet center of mass degrees of freedom parameterizing the motion of the M5-branes and the E-strings transform as a half hypermultiplet in the doublet representation of $SU(2)_L$, the factor appearing in the isometry
group $SO(4) \simeq (SU(2)_L \times SU(2)_R) / \mathbb{Z}_2$.

In what follows, we shall often make use of the following anomaly polynomials:
\begin{align}
\mathcal{I[}k\text{ M5's}]  &  =\frac{k^{3}}{24}\left(  c_{2}(L)-c_{2}%
(R)\right)  ^{2}-kI_{8}\\
\mathcal{I}\left[  k\text{ E-strings}\right]   &  =k^{3}\frac{\left(
c_{2}(L)-c_{2}(R)\right)  ^{2}}{6}+k^{2}\frac{(c_{2}(L)-c_{2}(R))I_{4}}%
{2}+k\left(  \frac{I_{4}^{2}}{2}-I_{8}\right)
\end{align}
The anomaly polynomial for M5-branes was obtained in reference \cite{Harvey:1998bx} and
that of the E-string theory was obtained in reference \cite{Ohmori:2014pca}. As in these references, we have
introduced the specific combinations:
\begin{align}
I_{4}  &  = - \frac{1}{2}(c_{2}(L)+c_{2}(R)) + \frac{1}{4}\left(  p_{1}(T)+\text{Tr}%
F_{E_{8}}^{2}\right) \\
I_{8}  &  =\frac{1}{48}\left(  (c_{2}(L)-c_{2}(R))^{2}+p_{2}(T)-\left(
c_{2}(R)+c_{2}(L)+\frac{1}{2}p_{1}(T)\right)  ^{2}\right).
\end{align}
We have also included the flavor symmetry $SU(2)_{L}$ present for both the E-string theory and the
M5-brane theory.

Now, an important subtlety in the F-theory realization of 6D\ SCFTs are curves
which support a singular fiber which does not have a corresponding gauge
group. Examples of this type include the configurations:%
\begin{equation}
\overset{I_{1}}{2},...,\overset{I_{1}}{2}\text{ \ \ and \ \ }\overset{II}{2}%
,...,\overset{II}{2}.
\end{equation}
Observe that this also occurs in the case of the example theory of line
(\ref{example theory}). The interpretation in F-theory of these cases was
discussed at length in \cite{Heckman:2015bfa}.\ As explained there, the key point is that
there will typically be some additional (half)\ hypermultiplets localized at
the intersection points of the discriminant locus. Since we will frequently
encounter situations like this in our analysis of anomaly polynomials, let us
pause to explain how to properly account for these contributions.

In the case of enhancement to an $I_{1}$ fiber, there is indeed a sense in
which we can interpret it as a single seven-brane wrapping the curve. That
means that morally speaking, there is a $u(1)$ gauge symmetry associated with
it. On the other hand, it is well-known that in any non-compact F-theory model
in six dimensions, such $u(1)$ factors are always coupled to background
two-form potentials, and are consequently eliminated by a generalized
St\"uckelberg mechanism. For the purposes of calculating an anomaly polynomial,
however, we can still formally view the $I_{1}$ fiber as contributing an
$su_{1}$ gauge symmetry. Factorization then proceeds as before.

In the case of an enhancement to a type $II$ fiber, we have a strongly coupled bound state of seven-branes of different
$(p,q)$ type. For this reason, it is simpler to just study the net change in the number of hypermultiplets connected
with the presence of such a fiber type.

\subsection{Examples with E-strings}

Let us illustrate these general considerations with a few examples. We mainly
focus on variations connected with the E-string theory as this is the most
common type of generalized matter we shall encounter. When a rank 1 E-string
connects two matter curves of gauge group $G_{L}$, $G_{R}$ i.e., when the
configuration takes the form
\[
...,G_{L},\underset{[G_{M}]}{1},G_{R},...
\]
a $G_{L}\times G_{R}$ subgroup of $E_{8}$ is gauged, leaving a leftover global
symmetry subgroup $G_{M}$, which may be trivial. This is accounted for by replacing
\begin{equation}
\text{Tr}F_{E_{8}}^{2}\rightarrow\text{Tr}F_{G_{L}}^{2}+\text{Tr}F_{G_{R}}%
^{2}+\text{Tr}F_{G_{M}}^{2}%
\end{equation}
Similarly, if the rank 1 E-string touches only a single curve carrying gauge
group $G$, i.e.
\[
\underset{\lbrack G_{M}]}{1},G,...
\]
one replaces
\begin{equation}
\text{Tr}F_{E_{8}}^{2}\rightarrow\text{Tr}F_{G}^{2}+\text{Tr}F_{G_{M}}^{2}%
\end{equation}
For the rank $k>1$ E-strings, the same decomposition of the flavor symmetry
also applies. However, the $c_{2}(L)$ dependence no longer drops out.  Note that it is possible to gauge the $SU(2)_L$ symmetry of the rank $2$ $E$-string, as discussed in \cite{Ohmori:2014kda}.  In the F-theory picture, this gauging corresponds to a configuration:
\[
1\,\,\overset{II}{2}\,\,\overset{IV^{ns}}{2}
\]

Next, we must add in the anomaly polynomial contributions from the
matter--namely those from any tensor multiplets, vector multiplets, or
hypermultiplets. The tensor multiplets each contribute
\begin{equation}
I_{\text{tensor}}=\frac{c_{2}(R)^{2}}{24}+\frac{c_{2}(R)p_{1}(T)}{48}%
+\frac{23p_{1}(T)^{2}-116p_{2}(T)}{5760}. \label{tensoreq}%
\end{equation}
The vectors of gauge symmetry $G$ contribute
\begin{equation}
I_{\text{vector}}=-\frac{\text{tr}_{\text{adj}}F^{4}+6c_{2}(R)\text{tr}%
_{\text{adj}}F^{2}+d_{G}c_{2}(R)^{2}}{24}-\frac{(\text{tr}_{\text{adj}}%
F^{2}+d_{G}c_{2}(R))p_{1}(T)}{48}-d_{G}\frac{7p_{1}(T)^{2}-4p_{2}(T)}{5760}.
\label{vectoreq}%
\end{equation}
Here, $d_{G}$ is the dimension of (the adjoint representation of) $G$.
Furthermore,
\begin{equation}
\text{tr}_{\text{adj}}F^{4}=t_{G}\text{tr}_{\text{fund}}F^{4}+\frac{3}{4}%
u_{G}(\text{Tr}F^{2})^{2}%
\end{equation}
and
\begin{equation}
\text{tr}_{\text{adj}}F^{2}=h_{G}^{\vee}\text{Tr}F^{2}.
\end{equation}
$h_{G}^{\vee}$ is the dual Coxeter number of $G$, and the values of the group
theory constants $t_{G}$, $u_{G}$ can be found in Appendix A of
\cite{Ohmori:2014kda}. Similarly, recall that the contribution from a hypermultiplet in
representation $\rho$ is
\begin{equation}
I_{\text{hyper}}=\frac{\text{tr}_{\rho}F^{4}}{24}+\frac{\text{tr}_{\rho}%
F^{2}p_{1}(T)}{48}+d_{\rho}\frac{7p_{1}(T)^{2}-4p_{2}(T)}{5760}.
\label{hypereq}%
\end{equation}
For fundamental hypermultiplets
\begin{equation}
\text{tr}_{\text{fund}}F^{4}=\frac{3}{4}x_{G}(\text{Tr}F^{2})^{2}%
\end{equation}
and
\begin{equation}
\text{tr}_{\text{fund}}F^{2}=s_{G}\text{Tr}F^{2}.
\end{equation}
The constants $x_{G}$ and $s_{G}$ can be found in Appendix A of
\cite{Ohmori:2014kda}. For hypermultiplets in other representations e.g.
spinor representations, these group theory factors can be read off from Table
2 of \cite{Grassi:2011hq}. Although it is technically a special case of
(\ref{hypereq}), it is useful to include the anomaly polynomial contribution
for a hypermultiplet in a mixed representation $(\rho,\mu)$:
\begin{align}
I_{\text{mixed}}  &  =\frac{d_{\mu}\text{tr}_{\rho}F_{\rho}^{4}+d_{\rho
}\text{tr}_{\mu}F_{\mu}^{4}}{24}+\frac{(\text{tr}_{\rho}F_{\rho}%
^{2})(\text{tr}_{\mu}F_{\mu}^{2})}{4}\\
&  +\frac{d_{\mu}(\text{tr}_{\rho}F_{\rho}^{2})p_{1}(T)+d_{\rho}%
(\text{tr}_{\mu}F_{\mu}^{2})p_{1}(T)}{48}+d_{\rho}d_{\mu}\frac{7p_{1}%
(T)^{2}-4p_{2}(T)}{5760}.
\end{align}

Finally, we must also add to the anomaly polynomial a Green-Schwarz term to
cancel the gauge anomaly. From the above, we see that the anomaly polynomial
may depend on the gauge field strength $F$ through either $\text{Tr}F^{4}$ or
$\text{Tr}F^{2}$. If there is an independent quartic Casimir invariant, the
former can never be a perfect square, and so factorization of the anomaly polynomial
is impossible unless all $\text{Tr}F^{4}$ terms drop out.
Fortunately, this is guaranteed to occur provided the anomaly cancellations
are imposed. As a result, for any 6D SCFT, it suffices to consider only the $\text{Tr}F^{2}$ dependence of
the anomaly polynomial. In general, the anomaly polynomial takes the form,
\begin{equation}
I_{\text{1-loop}}=a(\text{Tr}F^{2})^{2}+b(\text{Tr}F^{2})+c
\end{equation}
for some rational number $a$ and some functions $b$ and $c$ of $c_{2}(R)$,
$p_{1}(T)$, $p_{2}(T)$, and the field strengths of other gauge/flavor
symmetries. Now, adding in the Green-Schwarz term to cancel the gauge
dependent piece corresponds simply to completing the square in the variable
$\text{Tr}F^{2}$. We are left with,
\begin{equation}
I_{\text{tot}}=I_{\text{1-loop}}+I_{GS}=\underset{\text{1-loop}%
}{\underbrace{a(\text{Tr}F^{2})^{2}+b(\text{Tr}F^{2})+c}}%
-\underset{GS}{\underbrace{a(\text{Tr}F^{2}+\frac{b}{2a})^{2}}}=c-\frac{b^{2}%
}{4a}%
\end{equation}
If there are multiple gauge fields, the resulting expression $c-\frac{b^{2}%
}{4a}$ will also be a quadratic polynomial in the traces of their
squared-field strengths. As a result, we can simply repeat this procedure of
completing the square to remove all gauge field strength dependence, and the
result is the final expression for the anomaly polynomial of the theory.

\section{Examples of RG Flows\label{sec:EXAMPLES}}

In this section we apply the technology of the anomaly polynomials reviewed above to study whether
two theories are related by RG flow.  From the F-theory setup we know two theories
could potentially be related by RG flow if they
are connected by a change of parameters in the geometry.  But even in such
cases, it is often not very simple to determine if the two theories that
are connected by a change of parameters are related by RG flows or if they are dual
to one another and lead to equivalent conformal fixed points at the strong coupling point.
One outcome of analyzing a number of examples is that the anomaly polynomial
in all cases reveals an RG flow rather than a duality.
In fact these get rid of all the potential dualities among $(1,0)$ SCFTs that have distinct
weakly coupled tensor branches.  In other words, we find evidence that if two $(1,0)$
SCFTs in six dimensions have a different description on their respective tensor branches,
then they cannot be equivalent upon moving to the strong coupling point by tuning the vevs of the scalars in the tensor
multiplets. In particular this suggests that the classification of
reference \cite{Heckman:2015bfa}, which in principle only classified 6D theories up to dualities actually contains no redundancies, i.e.\
there are no non-trivial dualities.

Our plan in this section will be to calculate the anomaly polynomial for a UV SCFT, and study what becomes of this quantity
after applying various types of geometric deformations. In particular,
we check that when we restrict to the unbroken flavor symmetries common to the UV and IR
theories, the difference in the anomaly polynomial is a perfect square, in accord with the general considerations
of reference \cite{Intriligator:2014eaa}.

First, we study examples based on tensor branch flows. We then turn
to Higgs branch flows, both for weakly coupled systems as well as Higgsing by activating vevs for conformal matter.
Then, we check that the flow from the theory of two small $E_8$ instantons (with tensor branch $[E_8],1,2$)
to two single $E_8$ instanton theories (two independent $-1$ curves), is consistent with anomaly matching considerations.  We subsequently consider gauging the $E_8$ flavor symmetries of the above two systems, yielding the configurations 12,1,2 and 1,12,1, respectively.  From the perspective of geometry, these
two cases both lead to singular geometries which are connected by higher order deformations. In this case a direct computation of the
normalizability of these deformation moduli is quite non-trivial, but the analysis of the anomaly polynomials is again straightforward. Using
this method, we establish that there is indeed a flow from the $12,1,2$ theory to the $1,12,1$ theory. This shows that the picture
of RG flow is the correct one, and these theories are {\it not} dual.

We also consider the detailed examples of reference \cite{Heckman:2015bfa}, and the remarkable
correspondence between the discrete data of heterotic small instantons
probing an ADE\ singularity and the F-theory realizations of these SCFTs.
An important element of this match hinges on correctly identifying the
flavor symmetries on the two sides of the correspondence. Following the
general prescription outlined in reference \cite{Heckman:2015bfa}, there can
often be multiple F-theory geometries with the same flavor symmetries.
 Based on our discussion from the previous sections, it is natural to expect that
theories with the same flavor symmetry can actually be organized according to
a sequence of flows, as giving a vev to operators parameterizing the Higgs branch is orthogonal
to the data which fixes the asymptotic geometry of the instantons.
Indeed we find highly non-trivial evidence that $E_8$ instanton
theories with fixed bundle data organize themselves into classes uniquely labeled by asymptotic instanton
bundle data, the members of which are related to each other by RG flow.  The
generic theory with no tuned moduli (but the same flavor symmetries) is at the bottom of the flow. It would be interesting to
directly verify this picture on the (often strongly coupled) heterotic side.

\subsection{Flows on the Tensor Branch}

In this section we consider flows initiated by activating vevs for the scalars in a tensor multiplet, i.e.\ flows on the tensor branch
of the moduli space. Geometrically, this corresponds to resolving some of the $\mathbb{P}^1$'s of the base to finite size. Since all known
6D SCFTs are built out of tree-like structures, we can typically expect to get two or more decoupled SCFTs in the infrared. Our plan will be
to compute the anomaly polynomial for the UV theory, and to compare it with the anomaly polynomial of the full infrared theory. As expected
from reference \cite{Intriligator:2014eaa}, we find that when we restrict to the unbroken symmetry currents common to the UV and IR, that the
difference anomaly polynomials is a perfect square.

Consider for example a flow from a theory of the rank $k$ E-string theory to a lower rank E-string theory. In the UV theory, the tensor branch is given by $k$ $\mathbb{P}^1$'s intersecting as $\underset{k}{\underbrace{1,2,...,2}}$. We can initiate a flow to the infrared by choosing one of these $\mathbb{P}^1$'s, and resolving it to a large size, whilst still keeping the other curves collapsed at zero size. In the heterotic M-theory description, this amounts to pulling some number of M5-branes off the $E_8$ wall.

Our first item of business is therefore to determine the resulting IR theory. First of all, we have the rank $(k-m)$ E-string theory (i.e.\ the M5-branes still stuck on the wall). The flavor symmetries for this system are the $E_8$ wall of the heterotic nine-brane, and the isometries $SO(4) \simeq (SU(2)_L \times SU(2)_R) / \mathbb{Z}_2$ of the non-compact directions transverse to the M5-branes. We identify $SU(2)_R$ with the R-symmetry of our theory. Turning next to the $m$ M5-branes, we actually find two SCFTs. First, we have the interacting $A_m$ $(2,0)$ theory. Additionally, we have a single $(2,0)$ tensor multiplet which controls the center of mass degrees of freedom of the M5-brane. In terms of $(1,0)$ multiplets, this is a single hypermultiplet (i.e.\ the relative position of the M5-brane stack compared with the wall), and a $(1,0)$ tensor multiplet, the scalar of which controls the position of the M5-branes from the $E_8$ wall. For the M5-brane theory, the IR symmetry group is $SO(5)_R$, i.e.\ the directions transverse to the M5-brane, of which the $SO(4)$ isometries are common to both systems.

The case of pulling off a single M5-brane (i.e.\ $m = 1$) was obtained in \cite{Intriligator:2014eaa}, and our analysis amounts to a mild generalization of this result. Comparing the differences in the UV and IR anomaly polynomials, and restricting to the unbroken symmetry generators which are common to both systems, we find that the difference is indeed a perfect square:
\begin{equation}
\mathcal{I}_{UV} - \mathcal{I}_{IR} = \frac{m}{32} \Big( \mathrm{Tr}F_{E_8}^2 + (4k - 2m -2)c_2(L) - (4k-2m+2)c_2(R) + p_1(T) \Big)^2.
\end{equation}
Here, we have used the general expression for the anomaly polynomial of the E-string theory and the theory of M5-branes, which we reproduce below for the convenience of the reader:
\begin{align}
\mathcal{I[}k\text{ M5's}]  &  =\frac{k^{3}}{24}\left(  c_{2}(L)-c_{2}%
(R)\right)  ^{2}-kI_{8}\\
\mathcal{I}\left[  k\text{ E-strings}\right]   &  =k^{3}\frac{\left(
c_{2}(L)-c_{2}(R)\right)  ^{2}}{6}+k^{2}\frac{(c_{2}(L)-c_{2}(R))I_{4}}%
{2}+k\left(  \frac{I_{4}^{2}}{2}-I_{8}\right)
\end{align}
with notation as in section \ref{sec:APOLY}. We note that in the above expressions,
the center of mass degrees of freedom have been included. In the case of the tensor branch flow, the fact that the scalar of the tensor multiplet is neutral under all of the UV flavor symmetries means we can track anomaly matching for all of the terms in the IR as well. When we turn to Higgs branch flows, only a subset of global symmetries will be left unbroken.

As a more involved example, consider the case of M5-branes probing the geometry $\mathbb{R}_{\bot} \times \mathbb{C}^2 / \Gamma_{ADE}$, where
$\Gamma_{ADE}$ is a discrete ADE subgroup of $SU(2)$. The anomaly polynomial for this system was determined in reference \cite{Ohmori:2014kda}, and
we can also calculate it using the recursive method explained in that reference and reviewed in section \ref{sec:APOLY}. Letting $G$ denote the corresponding simply connected ADE Lie group associated with the singularity, we see that when we have $Q$ M5-branes,
the UV theory consists of a generalized quiver with $(G,G)$ conformal matter between each link. In the case of $Q$ M5-branes we have
$Q$ such links, and $(Q-1)$ generalized quiver nodes:
\begin{equation}
\text{UV Theory: }[G_L]  \oplus \underset{Q-1}{\underbrace{\overset{\mathfrak{g}}{(2)} \oplus \cdots \oplus \overset{\mathfrak{g}}{(2)}}} \oplus [G_R].
\label{UVtheory1}
\end{equation}
The anomaly polynomial for the UV SCFT is given by \cite{Ohmori:2014kda}:
\begin{equation}
\mathcal{I}[Q, \Gamma] = \mathcal{I}[Q\, \mathrm{M5's} \mathrm{\,at\,}\mathbb{C}^{2} / \Gamma] - I_{\mathrm{tensor}} - \frac{1}{2 Q} \left(\frac{1}{4} \mathrm{Tr}F^2_L - \frac{1}{4}\mathrm{Tr}F^2 \right)^2,
\end{equation}
where $\mathrm{Tr} F_L^2$ is associated with the flavor
symmetry $G_L$, with similar notation for $G_R$. Additionally,
the anomaly polynomial for the M5-brane probe theory is \cite{Ohmori:2014kda}:
\begin{equation}
\mathcal{I}[Q\,\mathrm{M5's} \mathrm{\,at\,}\mathbb{C}^{2} / \Gamma] = \frac{Q^3 \vert \Gamma \vert^2}{24} c_2(R)^2 - Q I_8 - \frac{Q \vert \Gamma \vert}{2} c_2(R) (J_{4,L} + J_{4,R}) - \frac{1}{2}I_L^{\mathrm{vec}} - \frac{1}{2} I_R^{\mathrm{vec}},
\end{equation}
where $J_{4,L}$ and $J_{4,R}$ are additional contributions from the orbifold singularity itself:
\begin{equation}
J_{4,L\,\mathrm{or}\,R} = \frac{\chi_{\Gamma}}{48} (4 c_2(R) + p_1(T)) + \mathrm{Tr}F^2_{L\,\mathrm{or}\,R}.
\end{equation}
with $\chi_{\Gamma} = r_\Gamma + 1 - \vert \Gamma \vert^{-1}$, and $r_\Gamma$ the rank of $G$.

\begin{figure}[ptb]
\begin{center}
                \includegraphics[trim=10mm 10mm 10mm 10mm, clip, width=50mm]{twoM5on.pdf}
                \includegraphics[trim=10mm 27mm 10mm 10mm, clip, width=50mm]{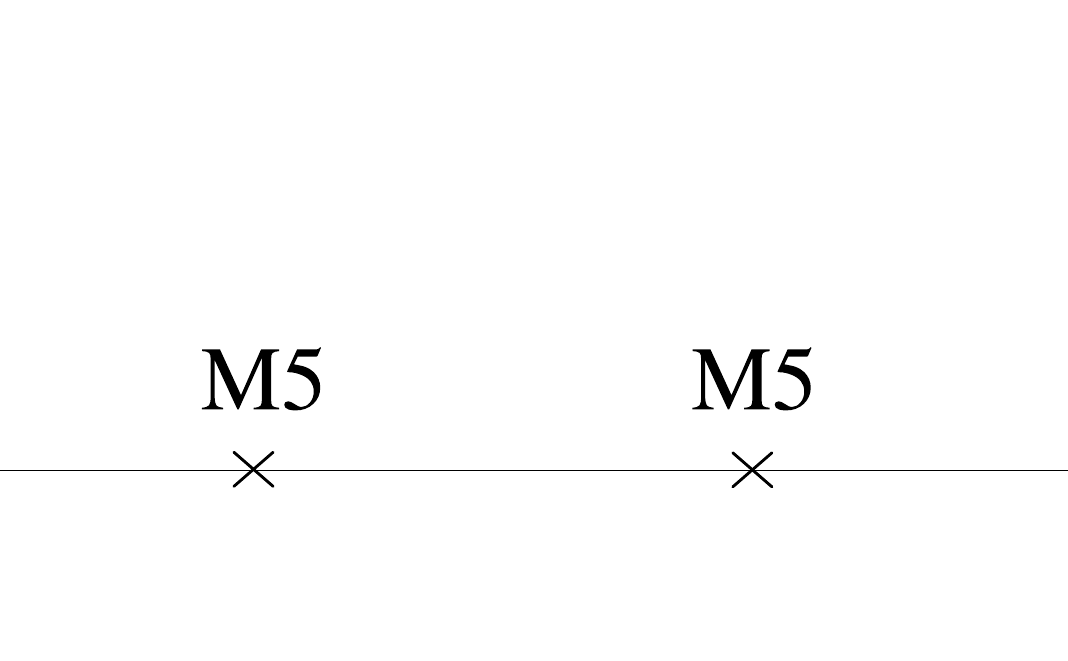}
\caption{Two M5-branes probing an $E_8$ singularity.  A tensor branch flow occurs when the two branes are separated along the singular locus.}
\label{twoM5tensor}
\end{center}
\end{figure}

Starting from this UV SCFT, we can initiate a tensor branch flow by sending one of our quiver nodes to weak coupling, i.e.\ by taking it to have very large size. In the M-theory description, this corresponds to keeping the M5-branes at the singularity, but partitioning them up as $Q = k + l$ into two separated stacks by moving them apart in the $\mathbb{R}_\bot$ direction. See figure \ref{twoM5tensor} for a depiction of this process for the case two M5-branes at an $E_8$ singularity.

The infrared SCFT actually consists of several decoupled SCFTs. First of all, we now have two decoupled generalized quivers. Additionally,
we have the volume of the curve we have blown up, and the associated free vector multiplets. Summarizing, here are the IR systems:
\begin{equation}
\text{IR Theory: }[G_L]  \oplus \underset{k-1}{\underbrace{\overset{\mathfrak{g}}{(2)} \oplus \cdots \oplus \overset{\mathfrak{g}}{(2)}}} \oplus [G_R^{\prime}] \decoup T_{(1,0)} \decoup V^{\oplus d_G} \decoup [G_L^{\prime}]  \oplus \underset{l-1}{\underbrace{\overset{\mathfrak{g}}{(2)} \oplus \cdots \oplus \overset{\mathfrak{g}}{(2)}}} \oplus [G_R]
\label{IRtheory1}
\end{equation}
where $V^{\oplus d_G}$ denotes the $d_G$ free vector multiplets obtained at zero gauge coupling.
Observe that the IR theory has additional flavor symmetries compared with the UV theory. In this case, the F-theory geometry allows us to cleanly
track these emergent symmetries. The anomaly polynomial for the IR system is:
\begin{equation}
\mathcal{I}_{IR} = \mathcal{I}[k , \Gamma] + \mathcal{I}[l , \Gamma] + \mathcal{I}[T_{(1,0)}] + \mathcal{I}[V^{\oplus d_G}].
\end{equation}

Let us now track the change in the anomaly polynomial for this system. Again, we need to make sure that if we restrict to global symmetries common to both systems that there is an exact factorization. So, we shall keep $\mathrm{Tr}F_L^2, \mathrm{Tr}F_R^2$. There is an additional $E_8$ global symmetry in the IR denoting the field strength of the quiver node we have decompactified, but we ignore this symmetry in what follows. For expository purposes we present the result for the case where there is no independent quartic Casimir invariant. This includes all of the exceptional gauge groups as well as $G = SU(2)$ and $G = SU(3)$.  The difference in the anomaly polynomials is then:
\begin{equation}
\mathcal{I}_{UV} - \mathcal{I}_{IR} = \frac{1}{32 k l Q} \left(-l \mathrm{Tr}F_L^2 - k \mathrm{Tr}F_R^2 
 + 2 k l Q \vert \Gamma \vert c_2(R) \right)^2
\end{equation}
which is manifestly a perfect square.

We emphasize that although we have used the M-theory description to guide the reader through the above examples, all of our computations can be phrased purely in terms of the F-theory realization of these SCFTs. As a more involved example along these lines, consider the F-theory model,
\begin{equation}
\text{UV Theory: } [SO(20)]\,\, \overset{\mathfrak{sp}_3}1 \,\, \overset{\mathfrak{su}_4}2 \,\, [SU(2)]
\end{equation}
The $SO(20)$ flavor symmetry comes from the 10 full hypermultiplets of $\mathfrak{sp}_3$ living on the $-1$ node, while the $SU(2)$ flavor symmetry comes from the two hypermultiplets of $\mathfrak{su}_4$ living on the $-2$ curve.  We claim this theory flows to four decoupled sectors in the IR,
\begin{equation}
\text{IR Theory: } [Sp(3)] \,\, \overset{\mathfrak{su}_4}2 \,\,[SU(2)] \decoup \mathcal{I}[T_{(1,0)}] \decoup  V^{\oplus 21} \decoup H^{\oplus 60}.
\label{IRsectors}
\end{equation}
Here, the $60$ free hypers come from the 10 \textbf{6}s of $\mathfrak{sp}_3$, while the $21$ free vectors come from the vector of $\mathfrak{sp}_3$ in the UV theory.  We know that the UV global symmetry $SO(20)$ and the IR global symmetry $Sp(3)$ are not preserved along the flow, so if we turn these symmetries off, we find that the difference between the anomaly polynomials is,
\begin{align}
\mathcal{I}_{UV}-\mathcal{I}_{IR}&=\frac{1}{64} (-48 c_2(R)+\text{Tr}F_{SU(2)}^2+2 p_1(T))^2.
\end{align}
This is once again a perfect square!  This provides a highly non-trivial check that the conformal sectors of the IR theory are indeed those shown in (\ref{IRsectors}).

\subsection{Flows on the Higgs Branch}

We can also consider RG flows initiated by complex structure deformations. At weak coupling, these are interpreted
as giving vevs to the scalars of a hypermultiplet, and at strong coupling these are interpreted as giving
vevs to conformal matter. In contrast to the case of tensor branch flows, the UV R-symmetry will be broken once we
activate such a vev. This means in particular that if we compare the anomaly polynomials in the UV and the IR, there is a priori
no reason for the contributions involving $c_2(R)$ to form a perfect square. This is simply because the UV and IR R-symmetries will generically
be different. Said differently, this is because the infrared R-symmetry is emergent. If, however, we restrict
to symmetries which are preserved along the flow (i.e.\ present in the UV and the IR), then we should still expect factorization
into a perfect square.

As a first class of examples, consider activating vevs for hypermultiplets. In F-theory this corresponds
to moving some seven-branes around.  In most cases, it is straightforward to identify flows in both physical terms
as well as in the F-theory geometry. Examples of this type include the vevs
for hypermultiplets of a gauge theory on the tensor branch. For example, in
the case of a $-2$ curve with an $I_{n}^{s}$ fiber, we have an $\mathfrak{su}%
_{n}$ gauge theory coupled to $2n$ flavors. The local presentation is that of
a curve of $A_{n+1}$ singularities which has been unfolded to a lower
singularity type:%
\begin{equation}
y^{2}=x^{2}+u^{2n}v^{n}\rightarrow x^{2}+u^{n}\underset{i=1}{\overset{n}{%
{\displaystyle\prod}
}}(uv+t_{i}), \label{weakexamp}%
\end{equation}
where $u=0$ corresponds to an $SU(2n)$ flavor symmetry, and $v=0$ corresponds
to our $\mathfrak{su}_{n}$ gauge symmetry.

There are, hower, other deformations of our geometry which do not obviously
correspond to deformations of the physical theory. Again focussing on the
above example, consider moving the marked points where the hypermultiplets are
localized. At generic values of the complex structure moduli, we can instead
consider the model:%
\begin{equation}
y^{2}=x^{2}+v^{n}\underset{i=1}{\overset{2n}{%
{\displaystyle\prod}
}}(u-u_{i}).
\end{equation}
This begs the question:\ do these moduli correspond to additional deformations
of our SCFT? For example, the presence of an $SU(2n)$ flavor symmetry has now
been obscured.

Now, in this special case, these additional moduli do not enter into the SCFT.
We can see this by explicitly calculating the anomaly polynomial in the tuned
case (all $u_{i}=0$) and the generic case as well. In both cases, we count
exactly $2n$ weakly coupled hypermultiplets, so anomaly matching
considerations indicate that there is no flow from the tuned case to the
generic case. Said differently, these two geometries lead us to the same 6D\ SCFT.

But in some cases a weakly coupled interpretation is unavailable, as will occur
any time we deal with conformal matter. As a simple example,
consider Higgsing a single conformal matter sector, say of $(E_8,E_8)$ type (Figure \ref{M5lifting}).
In the M-theory picture this corresponds to lifting off the M5-brane from the $E_8$ singularity locus.  The theory
flows to a theory of an extra hypermultiplet plus the degrees of freedom living
on the $E_8$ singularity.  The flavor symmetry, which was two copies of $E_8$, is now broken to the diagonal
$E_8$.   The degrees of freedom living on the $E_8$ singularity are given simply by the reduction from the 7D supersymmetric
$E_8$ gauge theory to a theory in 6 dimensions.  The vector symmetry is now a global symmetry, but there is also an adjoint hypermultiplet of $E_8$.
The anomaly polynomial for this diagonal $E_8$ is,
\begin{equation}
{\mathcal A}_{UV}-{\mathcal A}_{IR}={\mathcal A}_{[E_8]\oplus [E_8]}-{\mathcal A}_{(248+1){\rm Hyper}}=\frac{4163 c_2(R)^2}{8}-\frac{45}{2} c_2(R) \text{Tr}F^2-\frac{277 c_2(R) p_1(T)}{16}
,\end{equation}
Notice that again if we turn off $R$ we get a perfect match between the two anomaly polynomials.
Note that the coefficient of $c_2(R)^2$ has decreased, which for tensor branch flows follows from the
fact that the difference in the anomaly polynomial is a positive definite quadratic form.

\subsubsection{Instantons probing an $E_8$ Wall}

\begin{figure}
\begin{center}
\includegraphics[trim=10mm 30mm 30mm 14mm, clip, width=80mm]{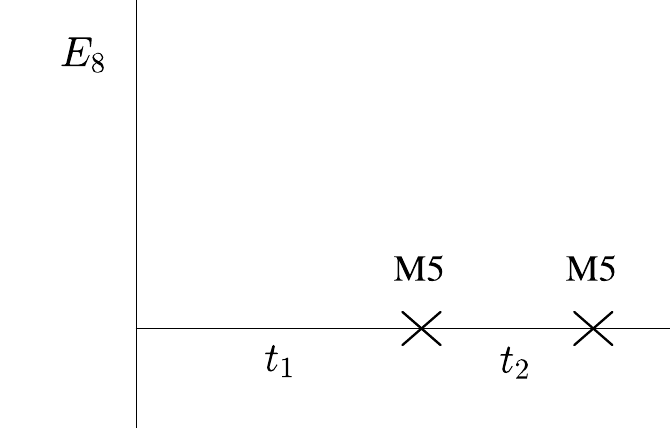}
\caption{M-theory realization of the theory of two heterotic small instantons.}
\label{twosmallinstantons}
\end{center}
\end{figure}

Consider the theory of two heterotic small instantons, which is realized in M-theory by two M5-branes probing an $E_8$ wall, as shown in Figure \ref{twosmallinstantons}.  In the F-theory picture, this corresponds to a base of the form
\begin{align}
[E_8] \,\, 1 \,\, 2
\end{align}
Here, the separation of the M5-branes corresponds to the sizes of the two $\mathbb{P}^1$s $t_1$, $t_2$, and the conformal limit is reached by taking the M5-branes to the wall, $t_1$, $t_2 \rightarrow 0$.  The $-1$ curve holds an $E_8$ flavor symmetry, corresponding to a non-compact locus of type $II^*$.

\begin{figure}
\begin{center}
\includegraphics[trim=10mm 20mm 30mm 10mm, clip, width=80mm]{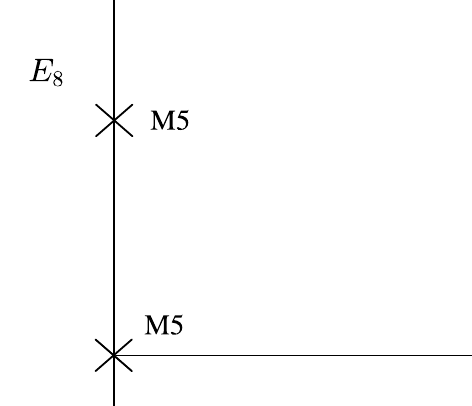}
\caption{M-theory realization of the theory of two heterotic small instantons.}
\label{twosmallinstantons2}
\end{center}
\end{figure}

Starting from this configuration with $t_1$, $t_2 \rightarrow 0$, we next consider separating the M5-branes along a direction parallel to the $E_8$ wall, as shown in Figure \ref{twosmallinstantons2}.  As the separation along the wall grows large, the theories living on the two $M_5$ branes decouple, and we are left with two decoupled theories of the form,
\begin{align}
[E_8] \,\, 1
\end{align}
In this manner, we get an RG flow from a rank-2 E-string theory to two decoupled rank-1 E-strings. In the UV theory, our global symmetry is $E_8 \times SU(2)_L \times SU(2)_R$. Now, when we move the instantons apart, there is a hypermultiplet controlling the relative positions of the two instantons. This specifies a vector in $SO(4)$ and as such breaks $SU(2)_L \times SU(2)_R$ to the diagonal subgroup $SU(2)_{diag}$. In the IR where the two instantons have decoupled, we find that each sector has (in addition to the common R-symmetry) its own $E_8 \times SU(2)_L$ flavor symmetry. In fact, in the special case of the one instanton theory, the $SU(2)_L$ acts only on the center of mass hypermultiplet of an E-string. This then is a simple example where the infrared R-symmetry is emergent.

The anomaly polynomial for the UV, rank two E-string theory is (including the center of mass degree of freedom):
\begin{align}
\mathcal{I}_{UV}&=\frac{31 c_2(R)^2}{12}-\frac{3}{4} c_2(R) \text{Tr}F_{E_8}^2-\frac{17 c_2(R) p_1(T)}{24}+\frac{1}{8} p_1(T) \text{Tr}F_{E_8}^2+\frac{(\text{Tr}F_{E_8}^2)^2}{16}\nonumber\\&+\frac{7 p_1(T)^2}{96}-\frac{p_2(T)}{24}.
\end{align}
Here, we have dropped the contribution from $c_2(L)$. Indeed, once we initiate a flow to the IR, we break to the diagonal of $SU(2)_L \times SU(2)_R$, and we only recover the R-symmetry in the IR. The anomaly polynomial for
the IR theory of two rank one E-strings (including each center of mass degreee of freedom) is:
\begin{align}
\mathcal{I}_{IR}&=\frac{13 c_2(R)^2}{12}-\frac{1}{4} c_2(R) \text{Tr}F_{(1)}^2-\frac{1}{4} c_2(R) \text{Tr}F_{(2)}^2-\frac{11 c_2(R) p_1(T)}{24}+\frac{1}{16} p_1(T) \text{Tr}F_{(1)}^2\nonumber \\&+\frac{(\text{Tr}F_{(1)}^2)^2}{32}+\frac{1}{16} p_1(T) \text{Tr}F_{(2)}^2+\frac{(\text{Tr}F_{(2)}^2)^2}{32}+\frac{7 p_1(T)^2}{96}-\frac{ p_2(T)}{24}
\end{align}
where again we have dropped the contribution from $c_2(L)$ from each of our E-string theories (as they are not common to the UV). We have,
however, kept the contribution from $c_2(R)$. Here, we have indicated the two $E_8$ flavor symmetries of the IR as $\mathrm{Tr}F^2_{(1)}$ and $\mathrm{Tr}F^{2}_{(2)}$. If we consider only the diagonal subgroup by taking $\text{Tr}F_{(1)}^2,\text{Tr}F_{(2)}^2 \rightarrow \text{Tr}F_{E_8}^2$ and setting the combination $\text{Tr}F_{(1)}^2-\text{Tr}F_{(2)}^2 $ to zero, we find,
\begin{align}
\mathcal{I}_{UV}-\mathcal{I}_{IR} = \frac{3 c_2(R)^2}{2}-\frac{1}{4} c_2(R) \text{Tr}F_{E_8}^2-\frac{c_2(R) p_1(T)}{4}
.
\end{align}
The terms depending solely on the field strength of diagonal $E_8$ subgroup have canceled!  Thus, we see that the diagonal $E_8$ subgroup has been preserved under the flow, whereas the $SU(2)_R$ symmetry has not.   Note again that the coefficient of $c_2(R)^2$ has decreased. Finally, we note that if we restrict to the diagonal $SU(2)_{diag} \subset SU(2)_L \times SU(2)_R$ common to both the UV and IR theories, the difference in the anomaly polynomials exactly vanishes.

Next, we consider compactifying the $E_8$ locus, resulting in an F-theory geometry with a $-12$ curve carrying a $II^*$ fiber.  Our new UV theory is specified by the base geometry,
\begin{align}
\overset{\mathfrak{e_8}}{12} \,\, 1 \,\, 2
\end{align}
The IR theory is specified by the geometry,
\begin{align}
1 \,\, \overset{\mathfrak{e_8}}{12} \,\, 1.
\end{align}
The Weierstrass model is:
\begin{equation}%
\begin{tabular}
[c]{|c|c|c|}\hline
& Equation & Curves\\\hline
UV & $y^{2}=x^{3}+v^{5}u^{2}+v^6$ & $(12),1,2$\\\hline
IR & $y^{2}=x^{3}+v^{5}
(u-u_{1})(u-u_{2}) +v^6.$ & $1,{(12)},1$\\\hline
\end{tabular}
\end{equation}
In the IR model, the elliptic model becomes too singular at the collision
of $v=0$ and $u=u_{i}$. This necessitates a blowup in the base, giving us a
$-1$ curve. In the tuned model, multiple blowups are required at the same
point, leading us to a model where the entire rank $2$ E-string theory touches
the $-12$ curve at a single point.

Based on our example from the previous subsection with weakly coupled hypermultiplets, we might at
first suppose that the moduli $u_{i}$ correspond to irrelevant deformations of
the physical theory. There are various ways to see, however, that the IR
model is actually distinct from the UV model.  In fact we can view this as gauging an
$E_8$ flavor symmetry of two theories related by RG flow, which we discussed in the previous
example.  Therefore this would suggest that the resulting theories should not be equivalent but
also related by RG flow (hence the labels ``UV" and ``IR").
 As this model is quite
singular, a direct mathematical analysis of the normalizability of the
deformations is rather challenging. In its place, to settle the matter, we can use our
analysis of anomaly polynomials to compare the two theories. First of all, we
can calculate the difference of the two anomaly polynomials. In the first
case, we have a rank $2$ E-string coupled to an $\mathfrak{e}_{8}$ gauge theory. Again,
we must properly take into account the contribution from the free
hypermultiplet describing the position of the E-string theory on the compact
curve. We compare this with the case of two individual E-strings. Here, we
have additional free hypermultiplets from moving each of the E-strings to a
new point on the $-12$ curve. Comparing the two anomaly polynomials, we have:%
\begin{align}
\mathcal{I}_{UV} - \mathcal{I}_{IR}=\frac{19}{4}c_2(R)^2 - \frac{c_2(R) p_1(T)}{20} . \label{anomdifference}%
\end{align}
Once again, we see that the terms proportional to $c_{2}(R)$ have
not canceled. This is a good indication that these theories have different
degrees of freedom and should be viewed as distinct.  Note that again the coefficient of
$c_2(R)^2$ has decreased.

Based on this, we conclude that the two theories are in fact distinct and
that the deformation moduli we have encountered should be viewed as operator
vevs which can move us between theories. Similar considerations hold for an $E_8$ gauge theory with rank $k$ E-strings
versus one with $k$ rank one E-strings. Indeed, we can then
label all the different possible theories according to the ways to
partition up $k$ points, i.e. Young tableaux with $k$
boxes.

We may perform a similar analysis for theories specified by F-theory geometries of the form,
\begin{align}
\overset{\mathfrak{g}}G,1,2\text{     \,\,\,\,    vs   \,\,\,\,    }1, \overset{\mathfrak{g}}G,1
\end{align}
For instance, consider
\begin{align}
\overset{\mathfrak{f_4}}5\,\,\underset{[G_2]}1\,\,2\text{     \,\,\,\,    vs   \,\,\,\,    }[G_2]\,\,1\,\, \overset{\mathfrak{f_4}}5\,\,1\,\,[G_2]
\end{align}
As before, we only consider the diagonal subgroup of the flavor symmetry $G_2 \times G_2$, as it is common to both the UV and the IR.
We find that the difference between the anomaly polynomials between the left (UV) and right (IR) theories is,
\begin{equation}
\mathcal{I}_{UV} - \mathcal{I}_{IR}=\frac{16 c_2(R)^2}{3}-\frac{5}{12} c_2(R) \text{Tr}F_{G_2}^2-\frac{c_2(R) p_1(T)}{6}
\end{equation}
Once again, the terms dependent only on the $G_2$ flavor symmetry have canceled whereas the ones dependent on $SU(2)_R$ have not, indicating that the $R$-symmetry has been broken along the flow while the diagonal $G_2$ flavor symmetry has been preserved.

From the F-theory perspective, this family of flows once again corresponds to separating the two marked points on the central $\mathbb{P}^1$ with
gauge group $G$.  The points mark the locations of the intersections with the two other $\mathbb{P}^1$s.  In the UV, these marked points lie on top of each other, but as we flow to the IR, the marked points are separated.

\subsubsection{Instantons on a $\Gamma_{E_{8}}$ Orbifold}

We next consider the case of heterotic
small instantons probing the orbifold singularity $\mathbb{C}^{2}%
/\Gamma_{E_{8}}$. As explained in \cite{DelZotto:2014hpa, Heckman:2015bfa}, these small instanton
theories have additional boundary data captured by a choice of homomorphism
$\Gamma_{E_{8}}\rightarrow E_{8}$. We focus on a particularly tractable
example where we take the $SU(2)$ subgroup of $(SU(2)\times E_{7})/%
\mathbb{Z}
_{2}\subset E_{8}$. This leads to an unbroken $E_{7}$ flavor symmetry. Thus,
we seek an F-theory model with an $E_{7}\times E_{8}$ flavor symmetry. Here,
the $E_{7}$ factor comes from the unbroken flavor symmetry of the
Horava-Witten wall, and the $E_{8}$ is from the orbifold singularity.

We find that there is actually more than one F-theory geometry with an
$E_{7}\times E_{8}$ flavor symmetry. On the tensor branch, these two theories
are:%
\begin{align}
\text{UV\ Model}  &  \text{: }[E_{7}]\oplus\overset{\mathfrak{e}_{7}}{8}%
\oplus\underset{k}{\underbrace{\overset{\mathfrak{e}_{8}}{(11)}\oplus\overset{\mathfrak{e}_{8}%
}{(12)}\oplus...\overset{\mathfrak{e}_{8}%
}{(12)}}}\oplus\lbrack E_{8}] \label{UVeq}\\
\text{IR Model}  &  \text{: }[E_{7}]\oplus\underset{k}{\underbrace{\overset{\mathfrak{e}_{8}}{(11)}\oplus\overset{\mathfrak{e}_{8}%
}{(12)}\oplus...\overset{\mathfrak{e}_{8}%
}{(12)}}}\oplus\lbrack E_{8}].
\label{IReq}
\end{align}
Here, we have used the compressed notation of reference \cite{Heckman:2015bfa} to indicate
various types of conformal matter:%
\begin{align}
E_{7}\oplus E_{7}  &  \simeq E_{7},1,2,3,2,1,E_{7}\\
E_{7}\oplus E_{8}  &  \simeq E_{7},1,2,3,1,5,1,3,2,2,1,E_{8}\\
E_{8}\oplus E_{8}  &  \simeq E_{8},1,2,2,3,1,5,1,3,2,2,1,E_{8}.
\end{align}

Our plan in this section will be to perform two computations. First of all, we
study the smoothing deformation where we give a vev to the $E_{7}\oplus E_{7}$
conformal matter. This can be interpreted as a standard Higgsing operation
whereby we flow to the diagonal $E_{7}$. Since one of our group factors is a
flavor symmetry, the resulting system again has an $E_{7}$ flavor symmetry,
i.e.\ we flow to the IR model. We shall then explicitly verify that there is
indeed a flow by comparing the values of the anomaly polynomials for these two theories.

So, let us begin with the analysis of the smoothing deformations of the
singularity. We begin with all curves blown down so that we just see the two
flavor branes for the leftmost $[E_{7}]$ and the rightmost $[E_{8}]$, which
collide along a complicated singularity. Denote the locus of the $E_{7}$
flavor brane by $\sigma=0$ and the $E_{8}$ flavor brane by $\tau=0$, so that
$\sigma^{3}\tau^{4}$ divides $f$ and $\sigma^{5}\tau^{5}$ divides $g$.

The family of singularities that includes both of these types is:%
\begin{equation}
y^{2}=x^{3}+\alpha\sigma^{3}\tau^{4}x+(\beta\tau+\gamma\sigma+\delta\tau
^{2})\sigma^{5}\tau^{5}.
\end{equation}
We first blow up the origin, focusing on the chart away from the $E_{8}$
flavor brane. To do this, we introduce new coordinates $(s,t)$ with
$\sigma=st$, $\tau=t$. Substituting and dividing by the appropriate powers of
$t$, we find the new equation:%
\begin{equation}
y^{2}=x^{3}+\alpha s^{3}t^{4}+(\beta+\gamma s+\delta t)s^{5}t^{5}.
\end{equation}
For simplicity, we set $\gamma=1$, giving the equation:%
\begin{equation}
y^{2}=x^{3}+\alpha s^{3}t^{4}+(\beta+s+\delta t)s^{5}t^{5}.
\end{equation}

When $\beta\neq0$, the additional small instanton is located at the extra zero
of $g$ along $t=0$, that is, at $s=-\beta$, $t=0$. We blow this up (for
arbitrary $\beta$). On the chart away from the $\mathfrak{e}_{8}$ gauge brane,
we can describe the blowup by two new coordinates $(u,v)$ satisfying
$s=uv-\beta$, $t=v$. Substituting, and dividing by appropriate powers of $v$,
we get:%
\begin{equation}
y^{2}=x^{3}+(uv-\beta)^{3}x+(u+\delta)(uv-\beta)^{5}.
\end{equation}
We see the $E_{7}$ flavor brane is at $uv=\beta$ for generic $\beta$. When
$\beta=0$, there are two $E_{7}$ branes: along the exceptional divisor $v=0$
we are getting an $\mathfrak{e}_{7}$ gauge brane, while along $u=0$ we get a
(noncompact) $E_{7}$ flavor brane. The physical interpretation of our
parameter $\beta$ is nothing but the vev of the recombination operator for
$(E_{7},E_{7})$ conformal matter \cite{Heckman:2014qba}. This is a good
indication that we have indeed initiated a flow from the UV to the IR.

To provide further evidence for this picture, we next consider the anomaly
polynomial for our two theories.  Here and henceforth, we restrict our anomaly polynomial computations to the case of a single $\mathfrak{e}_8$ gauge algebra, i.e.\ we consider the specific cases of (\ref{UVeq}) and (\ref{IReq}) with $k=1$.  The anomaly polynomial takes the form:
\begin{align}
\mathcal{I}_{UV} &=\frac{44297 c_2(R)^2}{2}-70 c_2(R) \text{Tr}F_L^2-\frac{237}{4} c_2(R) \text{Tr}F_R^2-\frac{451 c_2(R) p_1(T)}{4} \nonumber \\
 &+\frac{1}{16} \text{Tr}F_L^2 \text{Tr}F_R^2+\frac{5}{16} p_1(T) \text{Tr}F_L^2+\frac{5 (\text{Tr}F_L^2)^2}{32}+\frac{3}{8} p_1(T) \text{Tr}F_R^2+\frac{3 (\text{Tr}F_R^2)^2}{16} \nonumber \\ &+\frac{49 p_1(T)^2}{144}-\frac{7 p_2(T)}{36}
\end{align}
Here and in the following equations, $F_L$ is the field strength of the $E_7$ global symmetry on the left of the configuration, while $F_R$ is the field strength of the $E_8$ global symmetry on the right.  The anomaly polynomial for the IR theory is,
\begin{align}
\mathcal{I}_{IR} &=\frac{196523 c_2(R)^2}{24}-40 c_2(R) \text{Tr}F_L^2-\frac{165}{4} c_2(R) \text{Tr}F_R^2-\frac{3301 c_2(R) p_1(T)}{48} \nonumber \\
& +\frac{1}{16} \text{Tr}F_L^2 \text{Tr}F_R^2+\frac{5}{16} p_1(T) \text{Tr}F_L^2+\frac{5 (\text{Tr}F_L^2)^2}{32}+\frac{3}{8} p_1(T) \text{Tr}F_R^2+\frac{3 (\text{Tr}F_R^2)^2}{16}\nonumber \\
& +\frac{49 p_1(T)^2}{144}-\frac{7 p_2(T)}{36}
\end{align}
The difference is therefore,
\begin{align}
\mathcal{I}_{UV}- \mathcal{I}_{IR} &= c_2(R)  \left( \frac{335041 c_2(R)}{24}-30 \text{Tr}F_L^2-18  \text{Tr}F_R^2-\frac{2111  p_1(T)}{48} \right)
\end{align}
We note that again the difference vanishes when we set $R$ to zero and that again the coefficient of $c_2(R)^2$ has
decreased along the flow.  This holds true regardless of the number of $-12$ curves in the linear quiver, and the anomaly polynomials of the UV and IR theories in the flows studied in this paper will be equal up to terms proportional to $c_2(R)$.

\subsection{Further Examples of Instanton Flows}

Similar considerations apply for other flows involving heterotic small
instantons probing an orbifold singularity. Again, the picture of flows is
most transparent in the language of F-theory on the tensor branch. In this
subsection we collect some additional examples of this type, including the
relevant flow from UV to IR. \ For specificity, we again focus on the case
where the orbifold singularity is generated by $\Gamma_{E_{8}}$, the binary
octahedral group. In our conventions, the flavor $[E_{8}]$ is just to the
right of the first gauge $\mathfrak{e}_{8}$ curve and contracting the curves
in between the $\mathfrak{e}_{8}/E_{8}$ conformal matter leads to five
blowdowns on the gauged $\mathfrak{e}_{8}$ curve. We also blow up small
instantons on that $\mathfrak{e}_{8}$ curve. To keep our presentation somewhat
compact, we use the $\oplus$ notation to denote the curves generated by
performing a minimal number of blowups between two other gauge groups. Note
that this slightly generalizes the notion of \textquotedblleft
links\textquotedblright\ to the case where one of the gauge group factors is
not simply laced. Also, we note that there are well over one hundred different
small instanton probe theories which can be related by various flows. Our aim
here is therefore simply to select a few examples which illustrate the general
ideas presented earlier.
For each of these flows, we may calculate the change in matter charged under the (left, right) flavor symmetries between the UV and IR theories (but uncharged under any gauge symmetries) by comparing the relevant terms in the anomaly polynomial.  In particular
we find that the analog of an extra hypermultiplet that we got in the IR in previous examples, sometimes involves more
non-trivial matter representation of the flavor symmetry.  To bring this out, below when we mention
${\mathcal I}_{IR}$ we mean the ``na\"ive'' IR anomaly polynomial, which does not take into account this additional contribution from matter charged only under the global symmetries.  This difference in gauge-neutral matter appears in the tables below as $\varphi_{UV}-\varphi_{IR}$.\footnote{These contributions can be added back in to the anomaly polynomial by observing that there is no change in the coefficient of the $p_2(T)$ term in flowing from the UV to the IR.}

What follows is thus a small set of example flows which retain the same flavor
symmetry. In all of these examples, we let $F_L$ denote the field strength of the global symmetry on the left of the configuration, while $F_R$ is the field strength of the $E_8$ global symmetry on the right of the configuration.  Let us note that in accord with the classification results found in
reference \cite{Heckman:2015bfa}, there can sometimes be more than one UV
theory with the same left global symmetry. This is because there are different
group embeddings $\Gamma_{E_{8}}\rightarrow E_{8}$.

\begin{itemize}
\item $E_{7}\times E_{8}$ flavor symmetry%
\begin{equation}%
\begin{tabular}
[c]{|c|c|}\hline
$\text{Flow:}$ & $[E_{7}]\,\oplus\overset{\mathfrak{e_{7}}}{8}\oplus
\overset{\mathfrak{e_{8}}}{(12)}\,\,...[E_{8}]\rightarrow\lbrack
E_{7}]\,\oplus\overset{\mathfrak{e_{8}}}{(11)}\,\,...[E_{8}]$\\\hline
$\mathcal{I}_{UV}-\mathcal{I}_{IR}$ & $ \frac{335041 c_2(R)^2}{24}-30 c_2(R) \text{Tr}F_L^2-18 c_2(R) \text{Tr}F_R^2-\frac{2111 c_2(R) p_1(T)}{48}+\frac{7 p_1(T)^2}{5760}-\frac{p_2(T)}{1440} $\\\hline
$\varphi_{UV}-\varphi_{IR}$& $1 \times \textbf{(1,1)}$ \\\hline
\end{tabular}
\end{equation}

\item $F_{4}\times E_{8}$ flavor symmetry%
\begin{equation}%
\begin{tabular}
[c]{|c|c|}\hline
$\text{Flow:}$ & $[F_{4}]\,\oplus\overset{\mathfrak{e_{7}}}{8}\oplus
\overset{\mathfrak{e_{8}}}{(12)}\,\,...[E_{8}]\rightarrow\lbrack F_{4}%
]\oplus\overset{\mathfrak{e_{8}}}{(10)}\,\,...[E_{8}]$\\\hline
$\mathcal{I}_{UV}-\mathcal{I}_{IR}$ & $ -\frac{394637 c_2(R)^2}{24}+\frac{165}{4} c_2(R) \text{Tr}F_L^2+24 c_2(R) \text{Tr}F_R^2+\frac{2827 c_2(R) p_1(T)}{48}$ \\
&$-\frac{1}{16} p_1(T) \text{Tr}F_L^2-\frac{(\text{Tr}F_L^2)^2}{32}-\frac{203 p_1(T)^2}{5760}+\frac{29 p_2(T)}{1440} $ \\\hline
$\varphi_{UV}-\varphi_{IR}$& $1 \times \textbf{(26,1)}+ 3 \times \textbf{(1,1)}$ \\\hline
\end{tabular}
\end{equation}

\end{itemize}

\begin{itemize}

\item $SU(4)\times SU(2)$ flavor symmetry:

This is another case with a single flow:%
\begin{equation}%
\label{eq:su2xsu4}
\begin{tabular}
[c]{|c|c|}\hline
UV Theory
& $[SU(4)]\,\,\overset{I_{2}}{2}%
\,\,1\,\,\overset{[SU(2)]}{\overset{1}{\overset{\mathfrak{e_{7}}}{8}}%
}\,\,1\,\,\overset{\mathfrak{su_{2}}}{2}\,\,\overset{\mathfrak{g_{2}}%
}{3}\,\,1\,\,\overset{\mathfrak{f_{4}}}{5}\,\,1\,\,\overset{\mathfrak{g_{2}%
}}{3}\,\,{\overset{\mathfrak{su_{2}}}{2}}%
\,\,2\,\,1\,\,\overset{\mathfrak{e_{8}}}{(12)}\,\,...[E_{8}]$\\\hline
IR\ Theory & $[SU(4)]\,\,\overset{\mathfrak{su_{3}}}{2}%
\,\,\overset{\mathfrak{su_{2}}}{2}\,\,\overset{\mathfrak{su_{1}}%
}{2}\,\,1\,\,\underset{[SU(2)]}{\underset{2}{\underset{\mathfrak{su_{1}%
}}{\underset{1}{\overset{\mathfrak{e_{8}}}{(12)}}}}}\,\,...[E_{8}]$\\\hline
\end{tabular}
\end{equation}

The change in the anomaly polynomial is:%
\begin{align}
\mathcal{I}_{UV}
-\mathcal{I}_{IR}  & = \frac{354121 c_2(R)^2}{24}-\frac{149}{4} c_2(R) \text{Tr}F_L^2-\frac{\text{Tr}F_L^4}{96}-\frac{45}{2} c_2(R) \text{Tr}F_R^2 \nonumber \\
& -38 c_2(R) \text{Tr}F_T^2-\frac{2591 c_2(R) p_1(T)}{48}+\frac{1}{16} \text{Tr}F_L^2 \text{Tr}F_T^2+\frac{5}{96} p_1(T) \text{Tr}F_L^2\nonumber \\
& +\frac{(\text{Tr}F_L^2)^2}{32}+\frac{11}{192} p_1(T) \text{Tr}F_T^2+\frac{11 (\text{Tr}F_T^2)^2}{384}+\frac{161 p_1(T)^2}{5760}-\frac{23 p_2(T)}{1440}%
\end{align}
Here, $F_L$ is the field strength for the $SU(4)$ global symmetry, while $F_T$ is the field strength for the $SU(2)$ global symmetry.

For this flow, the global $SU(4)$ and $SU(2)_{\text{flavor}}$ symmetries are preserved, and we may calculate the change in matter between the
UV
and IR theories charged under the $SU(4) \times SU(2) \times E_8$ global symmetries.  The result is
\begin{equation}
\varphi_{UV} - \varphi_{IR} = 1 \times \textbf{(4,2,1)}+1 \times \textbf{(6,1,1)}+1 \times \textbf{(4,1,1)}+\frac{3}{2} \times \textbf{(1,2,1)}+2 \times \textbf{(1,1,1)}
\end{equation}
Recall that $\textbf{(6,1,1)}$ is the antisymmetric representation of $SU(4)$, and $ \textbf{(4,2,1)}$ is the bifundamental of $SU(4) \times SU(2)$.

\item $SU(6) \times E_8$ flavor symmetry:

In this case, it turns out that there is a sequence of flows we can do. The
main point is that we can also take non-minimal conformal matter, and this
sometimes comes with a flavor symmetry. So, taking this into account, we can
move around the associated additional E-string from a tuned point to a generic
point (i.e.\ the maneuver of moving marked points). Then, we can also activate
a vev for conformal matter, triggering a futher flow. We list the UV theory,
the intermediate theory, and the IR\ theory:

\begin{equation}%
\begin{tabular}
[c]{|c|c|}\hline
UV Theory & $[SU(6)]\,\,\overset{\mathfrak{su_{4}}}{2}%
\,\,{\overset{\mathfrak{su_{2}}}{2}}\,\,1\,\,\overset{\mathfrak{e_{7}}%
}{8}\oplus\overset{\mathfrak{e_{8}}}{(12)}\,\,...[E_{8}]$\\\hline
Intermediate & $[SU(6)]\,\,\overset{\mathfrak{su_{3}}}{2}%
\,\,1\,\,\overset{\mathfrak{e}_{6}}{6}\,\oplus\overset{\mathfrak{e_{8}}%
}{(12)}\,\,...[E_{8}]$\\\hline
IR\ Theory & $\lbrack SU(6)]\,\,\overset{\mathfrak{su_{3}%
}}{2}\,\,1\,\,\overset{\mathfrak{f_{4}}}{5}\,\,1\,\,\overset{\mathfrak{g_{2}%
}}{3}\,\,2\,\,2\,\,1\,\,\overset{\mathfrak{e_{8}}}{(11)}\,\,...[E_{8}%
]$\\\hline
\end{tabular}
\end{equation}

The change in the anomaly polynomial is:%
\begin{align}
\mathcal{I}_{UV}-\mathcal{I}_{\text{Intermediate}}  &= 5574c_2(R)^2-12c_2(R) \text{Tr}F_L^2+\frac{1}{96}\text{Tr}F_L^4-6c_2(R) \text{Tr}F_R^2 \nonumber \\
& -\frac{33 c_2(R) p_1(T)}{2} +\frac{1}{96} p_1(T) \text{Tr}F_L^2+\frac{7 p_1(T)^2}{960}-\frac{p_2(T)}{240} \\
\mathcal{I}_{\text{Intermediate}}-\mathcal{I}_{IR}& = \frac{189745 c_2(R)^2}{24}-18 c_2(R) \text{Tr}F_L^2-12 c_2(R) \text{Tr}F_R^2 \nonumber \\ & -\frac{1319 c_2(R) p_1(T)}{48}+\frac{7 p_1(T)^2}{5760}-\frac{p_2(T)}{1440}
\end{align}

The difference in gauge-neutral hypermultiplet content between the UV and intermediate theories can be read off from the anomaly polynomials: the UV theory has an extra fundamental of the $SU(6)$ flavor symmetry relative to the intermediate theory, while the intermediate theory has an extra neutral hyper relative to the IR theory.

\end{itemize}


\section{Conclusions and Future Directions \label{sec:CONC}}

Renormalization group flows from the UV to the IR are quite important in the study of quantum field theories.
In this paper we have taken some steps in establishing the structure of RG flows for
6D\ SCFTs which can be realized by F-theory compactification. By combining an
analysis of the change in the anomaly polynomial with the deformation theory
of the corresponding background Calabi-Yau threefold, we have determined
the structure of such flows. Let us briefly discuss some further
avenues of investigation.

The examples we have focussed on involve deformations which involve operator
vevs for 6D\ conformal matter. Using the geometric picture, we have seen (as
expected) that this conformal matter acts very much like a conventional
hypermultiplet. On the other hand, we have also observed some differences. For
example, moving the marked points where conformal matter is localized can
trigger a flow, which is different from the case of conventional matter. It
would be quite interesting to develop further details of this structure, and
the corresponding flows.

In this note our primary emphasis was on giving examples of the kinds of
phenomena one can expect to encounter in the study of 6D\ flows. Given the
fact that there is also a concrete classification of 6D\ SCFTs generated by
F-theory available, it would be natural to map out all possible flows between
these theories.

Finally, our analysis here has focused on general questions connected with
RG\ flow. There is also a holographic interpretation of RG flow. It would be
interesting to consider those 6D\ SCFTs with a holographic dual and determine
the detailed structure of such flows and how they interpolate between
different gravity duals.

\section*{Acknowledgements}

We thank K. Yonekura for several helpful discussions and correspondence. The
authors also thank the organizers of the conference on physics and geometry of
F-theory held at the Max Planck Institute for Physics in\ Munich for
hospitality, where some of this work was completed. The work of DRM is
supported by NSF grant PHY-1307513. The work of TR and CV is supported by NSF
grant PHY-1067976. TR is also supported by the NSF GRF under DGE-1144152.




\bibliographystyle{utphys}
\bibliography{sixDflows}

\end{document}